\documentclass[prd,nofootinbib,tightenlines,superscriptaddress,showpacs,preprintnumbers]{revtex4}

\usepackage{amsmath,amssymb,amsfonts,graphicx,subfigure,wrapfig,slashed}

\newcommand{\vect}[1]{\boldsymbol{#1}}

\begin{document}

\preprint{JLAB-THY-09-1002,~NT@UW-09-12}

\title{The NJL-jet model for quark fragmentation functions}

\author{T. Ito}
\email[Corresponding author:~]{8atad002@mail.tokai-u.jp}
\affiliation{Department of Physics, School of Science, Tokai University,
             Hiratsuka-shi, Kanagawa 259-1292, Japan}
\author{W. Bentz}
\email{bentz@keyaki.cc.u-tokai.ac.jp}
\affiliation{Department of Physics, School of Science, Tokai University,
             Hiratsuka-shi, Kanagawa 259-1292, Japan}
\author{I.~C.~Clo\"et}
\email{icloet@phys.washington.edu}
\affiliation{Department of Physics, University of Washington, Seattle, WA 98195-1560, USA}
\author{A.~W.~Thomas}
\email{awthomas@jlab.org}
\affiliation{Jefferson Lab, 12000 Jefferson Avenue, Newport News, VA 23606, USA and \\
College of William and Mary, Williamsburg, VA 23187, USA}
\author{K. Yazaki}
\email{yazaki@phys.s.u-tokyo.ac.jp}
\affiliation{Radiation Laboratory, Nishina Accelerator Research Center,
RIKEN, Wako, Saitama 351-0198, Japan}

\begin{abstract}
A description of fragmentation functions which satisfy the momentum and
isospin sum rules is presented in an effective quark theory. Concentrating
on the pion fragmentation function, we first explain why the 
elementary (lowest order) fragmentation process $q \rightarrow q \pi$
is completely inadequate to describe the empirical data, although the
``crossed'' process $\pi \rightarrow q \bar{q}$ describes the
quark distribution functions in the pion reasonably well. Taking into
account cascade-like processes in a generalized jet-model
approach, we then show that the momentum and isospin sum rules can be satisfied
naturally, without the introduction of ad hoc parameters. 
We present results for the Nambu--Jona-Lasinio (NJL) model in the invariant 
mass regularization scheme and compare them with the empirical parametrizations. 
We argue that the NJL-jet model, developed herein, provides a useful framework with which to 
calculate the fragmentation functions in an effective chiral quark theory.  
\end{abstract}


\maketitle

\vspace{-1em}
\section{Introduction}

Quark distribution and fragmentation functions are the basic nonperturbative
ingredients for a QCD-based analysis of hard scattering 
processes~\cite{Field:1976ve,Altarelli:1979kv,Collins:1981uw,Jaffe:1996zw,
Ellis:1991qj,Barone:2001sp}. 
Distribution functions can be extracted by analyzing inclusive 
processes~\cite{Martin:2003sk,Sutton:1991ay} and their
description in terms of effective quark theories of QCD has been quite
successful~\cite{Cloet:2005pp,Wakamatsu:1997en}. In recent years there has been 
a significant effort to extract the fragmentation functions by analyzing 
inclusive hadron production (semi-inclusive)
processes in $e^+\,e^-$ annihilation, deep-inelastic lepton-nucleon
scattering and proton-proton collisions~\cite{Hirai:2007cx,deFlorian:2007aj}.
Besides being of fundamental interest in their own right, 
knowledge of fragmentation functions is essential for the extraction of the transversity 
quark distribution functions~\cite{Ralston:1979ys,Barone:2001sp} 
from data, and to analyse several other interesting effects in semi-inclusive
processes~\cite{Sivers:1989cc}. 

Because of the importance of the fragmentation functions many attempts 
have been made to describe them using effective quark theories~\cite{Londergan:1996vf}. 
However, in order to achieve reasonable agreement with the 
empirical parametrizations it was necessary to introduce 
new parameters, like normalization constants, which cannot be 
justified on theoretical grounds. A description of fragmentation
functions within effective quark theories, which automatically satisfies the
relevant sum rules~\cite{Collins:1981uw} and describes the empirical data 
reasonably well -- without introducing new parameters into the theory --
has hitherto not been achieved. 

This failure to describe the fragmentation functions in the same framework
which is successful at describing the distribution functions is surprising, because
there exists a general relation, the so called Drell-Levy-Yan (DLY) relation
\cite{Drell:1969jm,Blumlein:2000wh},
which suggests a way to compute the fragmentation functions by analytic 
continuation of the distribution functions into the region of 
Bjorken $x>1$. Although the derivation of this relation appears to be
very general (as we show in Appendix A), 
the basic \textit{assumption} that the distributions and 
fragmentations are essentially one and the same function, 
defined in different regions of the scaling variable, has not been proven. 
Moreover, the approximations used to calculate the distribution functions 
may not be sensible for the
fragmentation functions and vice versa. For example, in the 
fragmentation process of a quark into a pion, $q \rightarrow \pi + n$, 
where $n$ is a spectator,
there is no \textit{a priori} reason to truncate $n$ to a single quark state, as the
DLY crossing arguments would suggest for the case of a Bethe-Salpeter type
vertex function for $\pi \rightarrow q \bar{q}$. 
One can actually give a quantitative argument that the lightest 
component of $n$ is dominant only if the scaling variable $z$ is very 
close to unity~\cite{Boros:1999zc}.  
   
On the other hand, the phenomenological quark jet-model, 
as formulated originally by Field and Feynman~\cite{Field:1977fa}, 
suggests that the meson observed in a semi-inclusive process is one 
among many, that is, the spectator state $n$ contains many
mesons. This model is based on a \textit{product ansatz} for a chain of
elementary fragmentation processes, where 
in each step a certain fraction of the quark momentum is transferred to 
a meson, until eventually a very soft quark remains. This final quark is assumed to
annihilate with the other remnants of the process without producing 
further observable mesons.\footnote{This picture of independent fragmentation 
is appealing because of its simplicity. More elaborate models for hadronization 
are the string model \cite{Andersson:1985qr} or the cluster model \cite{Field:1982dg}, which
are suitable for Monte Carlo analysis.} 
In order for all of the quark light-cone
momentum to be transferred to the mesons, it is actually necessary to
assume an \textit{infinite} number of steps (mesons) in the decay chain, as will
be explained in more detail in Section~\ref{sec:product}. In this case,
it is possible to satisfy the momentum sum rule for fragmentation
functions~\cite{Collins:1981uw}, which is assumed valid in QCD-based fits to the
data~\cite{Hirai:2007cx,deFlorian:2007aj}. Clearly, this sum rule cannot be satisfied in a 
single step elementary fragmentation process.

The purpose of this paper is to apply the method of the quark jet-model
to calculate the spin-independent fragmentation functions in an effective 
chiral quark theory, which has proven to be very successful for the description 
of quark distribution functions~\cite{Cloet:2005pp,Bentz:1999gx,Cloet:2007em}. We will concentrate on quark fragmentation
into pions within the Nambu--Jona-Lasinio (NJL) model~\cite{Nambu:1961tp}, 
however the methods illustrated
here can easily be extended to other fragmentation channels 
and applied within other effective quark theories. 
In order to reconcile the quark jet-model with our present 
NJL model description, we will introduce a generalized product ansatz, 
which allows for the fragmentation of a quark into a finite number of pions 
according to a certain distribution function, and in the end we take the limit 
of infinitely many pions. We will show how the momentum and
isospin sum rules emerge naturally without introducing any
new parameters into the theory. Our numerical results will demonstrate that
this NJL-jet model provides a very 
reasonable framework for describing the fragmentation functions.     

This paper is organized as follows: In Section~\ref{sec:operator} we 
begin with the operator definitions for the quark distribution and fragmentation 
functions and move on to discuss the sum rules and the DLY relation. 
In Section~\ref{sec:elementary}
we give the expressions for the elementary fragmentation functions
in the NJL model and discuss their physical interpretations and 
sum rules. In Section~\ref{sec:product}
we introduce the generalized product ansatz to describe a chain of
elementary fragmentation 
processes in the spirit of the quark jet-model, derive the integral equation
for the fragmentation function and discuss the momentum and isospin 
sum rules. In Section~\ref{sec:results}
we explain the model framework for the numerical calculations, present
results and compare them with the empirical fragmentation functions.
A summary is given in Section~\ref{sec:summary}.

\section{Operator definitions and sum rules}
\label{sec:operator}

Operator definitions and sum rules for fragmentation functions were first given 
in Ref.~\cite{Collins:1981uw} and were further elucidated in Ref.~\cite{Collins:1992kk}. 
In this Section we summarize the basic relations for the fragmentation functions 
and for clarity include those for the distribution functions also.
The spin-independent distribution function of a quark of flavour $q$ inside 
a hadron of spin-flavour $h$ (for example $h=p\!\uparrow, \pi^+$, etc.)
and the spin-independent fragmentation function for $q \rightarrow h$ 
are defined by
\begin{align}
f_q^h(x) &= \frac{1}{2} \int \frac{d\omega^-}{2\pi} e^{ip_- \omega^- x}
\,\hat{\sum}_n \langle p(h)|\overline{\psi}(0)|p_n \rangle\, \gamma^+\,
\langle p_n|\psi(\omega^-)|p(h) \rangle,    
\label{fdef} \\
D_q^h(z) &= \frac{z}{12} \int \frac{d\omega^-}{2\pi} e^{i p_- \omega^-/z}
\,\hat{\sum}_n \langle p(h),{p_n} | \overline{\psi}(0)|0 \rangle\, \gamma^+\,
\langle 0| \psi(\omega^-)| p(h),{p_n} \rangle.
\label{ddef}
\end{align}
The field operators
refer to a quark of flavour $q$, although it is not indicated explicitly.
The symbol $p(h)$ refers to a hadron $h$ with momentum $p$ and $p_n$
labels the spectator state. 
The light-cone components of a 4-vector are defined as 
$a^{\mu}=(a^+, a^-, {\vect a}_T)$ with $a^{\pm} = (a^0 \pm a^3)/\sqrt{2}$.
Covariant normalization is used throughout this 
paper and the summation symbol $\hat{\sum}_n$ includes
an integration over the on-shell momenta $p_n$.\footnote{In this normalization 
$\langle p'(h')|p(h)\rangle =2 p_- (2\pi)^3 \delta^{(3)}\left({\vect p}'- {\vect p}\right) 
\delta_{hh'}$ and
$|p(h), {p_n} \rangle = \sqrt{2 (2\pi)^3 p_-} \, a^{\dagger}_h(p) 
|{p_n} \rangle$, with $\left[a_h(p'),a_h(p)\right]_{\pm} = 
\delta^{(3)}\left({\vect p}'-{\vect p}\right)$. 
The summation defined by
$\hat{\sum}_n \equiv \sum_n \int \frac{d^4 p_n}{(2 \pi)^3}\
\delta\!\left( p_n^2 - M_n^2 \right)\,\Theta(p_{n0})$, where $M_n$ is
the invariant mass of $n$,
can also be expressed in terms of light-cone variables.
\label{foot:2}} 
Both expressions in Eqs.~\eqref{fdef} and \eqref{ddef} refer to a frame where $\vect{p}_T=0$.
The physical content of the functions in Eqs.~\eqref{fdef} and \eqref{ddef} is most
transparent if we introduce the ``good'' light-cone quark field 
$\psi_+$~\cite{Jaffe:1991ra,Burkardt:1995ct,Bentz:1999gx}, which is
defined by $\psi_+ \equiv \Lambda_+ \psi$ where $\Lambda_+ = \frac{1}{2} \gamma^- \gamma^+$
and can be expressed as the Fourier decomposition
\begin{align}
\psi_+(\omega^-) = \int_0^{\infty}\frac{dk_-}{\sqrt{2 k_-}} 
\int \frac{d^2 k_T}{(2\pi)^{3/2}}
\sum_{\alpha} \left(b_{\alpha}(k)\, u_{+ \alpha}(k)\, e^{-i k_- \omega^-}
+ d^{\dagger}_{\alpha}(k)\, v_{+ \alpha}(k)\, e^{i k_- \omega^-}\right). 
\label{eq:fourier} 
\end{align}
The index $\alpha$ denotes the spin-color of a quark with flavour $q$ 
and the spinors are normalized as 
$u^{\dagger}_{+ \alpha'}(k)\, u_{+ \alpha}(k)=v^{\dagger}_{+ \alpha'}(k)\, 
v_{+ \alpha}(k)=\sqrt{2}\, k_-\, \delta_{\alpha',\alpha}$. Substituting these expressions
into  Eqs.~\eqref{fdef} and \eqref{ddef}
and using the result $\overline{\psi} \gamma^+ \psi = \sqrt{2} \psi_{+}^{\dagger} \psi_{+}$,
%
gives the following relations which are independent of the normalization 
of states:
\begin{align}
f_q^h(x)\, dx &= dk_- \int d^2k_T \sum_{\alpha} 
\frac{\langle p(h)| b^{\dagger}_{\alpha}(k) b_{\alpha}(k)|p(h) \rangle} 
{\langle p(h)|p(h) \rangle },  
\label{if} \\
D_q^h(z)\, dz &= \frac{z^2}{6} dp_- \int d^2k_T
\sum_{\alpha} \frac{ \langle k(\alpha)|a^{\dagger}_h(p) a_h(p)| 
k(\alpha) \rangle} {\langle k(\alpha)| k(\alpha) \rangle }. 
\label{id}
\end{align}
Here $dx=dk_-/p_-$, that is, $k_-=x\,p_-$ for some fixed $p_- >0$
and $dz = dp_-/k_-$, implying $p_-=z\,k_-$ for some fixed $k_- >0$. 
The creation and annihilation operators, $a^{\dagger}_h$ and $a_h$, refer to the 
hadron $h$ (see footnote~\ref{foot:2}) and $k(\alpha)$ labels a  quark state
of flavour $q$ with momentum $k$ and spin-color $\alpha$.

According to Eq.~\eqref{if} we can interpret $f_q^h(x)$ 
as the light-cone momentum distribution of $q$ in $h$, where a sum 
over the spin-color of $q$ is understood, while the spin of $h$ is fixed. 
However, the result is independent of this spin direction, since we will 
only consider the spin-independent distributions. 
As mentioned earlier, Eq.~\eqref{id} 
refers to the frame where the produced hadron $h$ has $\vect{p}_{T}=0$, but the
fragmenting quark has non-zero $\vect{k}_T$. To interpret this result
as a distribution of $h$ in $q$, it is necessary to make a Lorentz transformation
to the frame where $\vect{k}_{\perp}=0$, but $h$ has non-zero $\vect{p}_{\perp}$
(note the distinction between the subscripts $T$ and ${\perp}$). 
This is discussed in detail in Refs.~\cite{Collins:1981uw,Barone:2001sp}, with the 
result that one can simply substitute
\begin{align}  
\vect{k}_T = - \frac{{\vect p}_{\perp}}{z},  
\label{sub}
\end{align}
leaving everything else unchanged. We then obtain from Eq.~\eqref{id} the result
\begin{align}
D_q^h(z)\, dz = \frac{1}{6} dp_- \int d^2 p_{\perp}
\sum_{\alpha} \frac{ \langle k(\alpha)|a^{\dagger}_h(p) a_h(p)| 
k(\alpha) \rangle} {\langle k(\alpha)| k(\alpha) \rangle }, 
\label{id1}
\end{align}
where the fragmenting quark now has $\vect{k}_{\perp}=0$.
According to Eq.~\eqref{id1} we can interpret $D_q^h(z)$ as the
light-cone momentum distribution of $h$ in $q$, where the factor 1/6
indicates an \textit{average}~\cite{Collins:1992kk} of the
spin-color of $q$,  while the spin of $h$ is fixed.\footnote{For the generalized case 
where $h$ can also be a quark, we summarize the definitions as follows: 
$f_q^h(x)$ refers to fixed flavours of $q$ and $h$, while all other quantum numbers of 
$q$ (spin, color, etc) are summed over, with those of $h$ are fixed.
$D_q^h(z)$ refers to fixed flavours of $q$ and $h$, with an \textit{average}
over the other quantum numbers of $q$ (spin, color, etc), while those
of $h$ are fixed. This definition has the advantage that in a semi-inclusive
process, which involves the product $f_q^T(x) D_q^h(z)$, the quark 
spin-color
summation is included in $f$ but not in $D$, which avoids double
counting.
\label{foot:3a}
}
In fact, for the elementary distribution and fragmentation functions 
considered in the next section, the naively expected relation
\begin{align} 
D_q^h(z) = \frac{1}{d_h}\,f_h^q(z),
\label{rel}
\end{align} 
is valid. Where $d_h$ is the spin degeneracy, or, in the general case, 
the spin-color degeneracy of $h$. Generally however, this relation is not
necessarily valid, because $q$ is off-shell (its virtuality being 
determined kinematically by the scaling variable and the transverse momentum) 
and $h$ is on-shell, which
breaks the naive symmetry under the interchange $q \leftrightarrow h$.  

To obtain the momentum sum rule from 
Eq.~\eqref{id1} we multiply both sides by $z=p_-/k_-$,   
integrate over $z$ from 0 to 1 and sum over
$h$.\footnote{A subtle point here is that in order to get an integral 
$\int_0^{\infty}
d p_-$ on the right hand side of Eq.~\eqref{p}, one has to choose 
$k_- = \infty$. This 
does not influence the result, which depends only on $z$.}
Then one notes that the momentum operator, represented in terms of hadron 
operators, is given by
\begin{align}
{\hat P}_- \equiv \sum_h \int_0^{\infty} dp_- \int d^2 p_{\perp} 
\left(p_-\, a^{\dagger}_h(p)\, a_h(p) \right).   
\label{p}
\end{align}
By assuming that the quark state $|k(\alpha)\rangle$ in Eq.~\eqref{id1} is an 
eigenstate of this operator with eigenvalue $k_-$, we obtain the momentum sum rule
\begin{align}
\sum_h \int_0^1 dz\,z\, D_q^h(z) = 1.    
\label{mom}
\end{align}
The physical content of Eq.~\eqref{mom} is that $100\%$ of the initial quark
light-cone momentum ($k_-$) is transferred to the hadrons. The 
condition which lies at the basis of Eq.~\eqref{mom} is that the initial 
quark state is an eigenstate of the momentum operator, Eq.~\eqref{p}, expressed 
solely in terms of hadrons. That is, the quark hadronizes completely in the sense
that it gives all of its light-cone momentum to the hadrons.  

A similar argument leads to the isospin sum rule~\cite{Collins:1981uw}, namely
\begin{align}
\sum_h \int_0^1  dz \, t_h \, D_q^h(z) = t_q,    
\label{iso}
\end{align}
where $t_q$ and $t_h$ denote the 3-components of the isospins of $q$ and $h$.
The physical content of this sum rule is that all of the isospin of the
initial quark is transferred to hadrons, 
which is possible since the definition in Eq.~\eqref{ddef} implies an average over
the isospin of the soft quark remainder of a fragmentation
chain (see Section~\ref{sec:product}). In general, there is no sum rule for the
baryon number or electric charge, because the baryon number or average
electric charge of the quark remainder is not zero.\footnote{The average 
of the electric charge is zero only if SU(3) flavour symmetry is assumed.}
If we simply integrate both sides of Eq.~\eqref{id1} over $z$, we get the
hadron multiplicity, which can be interpreted as the \textit{number of mesons per quark}.
However, there is no conservation law which leads to a sum rule for
the multiplicity.

There is an interesting relation based on charge conjugation
and crossing symmetry, between the fragmentation function for physical $z<1$ 
and the distribution function for unphysical $x>1$:
\begin{align}
D_q^h(z) = (-1)^{2(s_q + s_h) + 1} \, \frac{z}{d_q}\, 
f_q^h\left(x=\frac{1}{z}\right),  
\label{dly}
\end{align}
which is called the Drell-Levy-Yan (DLY) relation~\cite{Drell:1969jm,Blumlein:2000wh}. 
Here $s_q$ and $s_h$ are the spins of $q$ and $h$ respectively, and
$d_q$ is the spin-color degeneracy of $q$. 
We derive this relation using two independent methods in Appendix A. 
The first approach, which follows
the original arguments~\cite{Drell:1969jm}, compares the hadronic tensors
for $e h \rightarrow e' X$ (inclusive DIS) and $e^+ e^- \rightarrow h X$
(inclusive hadron production), and uses crossing relations for matrix
elements of the current operator. The second method -- which to the best of
our knowledge has not been published before -- starts directly from
the operator definitions in Eqs.~\eqref{fdef} and \eqref{ddef} and uses charge
conjugation and crossing symmetries for matrix elements of the quark
field operator.
If one has an effective quark theory to calculate the quark distribution
functions, Eq.~\eqref{dly} would suggest a straightforward way to
obtain the fragmentation functions. However, as will become clear in the following
sections, for the lowest order (elementary) processes
such an attempt leads to disastrous results. That is, the fragmentation functions
obtained in this way are one or two orders of magnitude smaller than the
empirical functions and the sum rules in Eqs.~\eqref{mom} and \eqref{iso} are
not satisfied.

The reasons why Eq.~\eqref{dly} fails in actual applications are
as follows: (i) It is based on the \textit{assumption} that the distribution
functions can be continued analytically beyond $x=1$. However, it is
well known that the $Q^2$ evolution 
equations lead to singularities at $x=1$,
which are (regularized) infrared singularities arising from the vanishing
gluon mass~\cite{Ellis:1991qj,Altarelli:1977zs}. These render an analytic 
continuation impossible. 
Someone may still argue that Eq.~\eqref{dly} should be
used only at the low energy (model) scale, however it is actually broken
there also, because of the cut-off regularization. We will discuss
this point in detail in the next section.
(ii) Most importantly, approximations which work
reasonably well for the distribution functions may not be sensible for the
fragmentation functions and vice versa. 
For example, the assumption that the pion 
is a $q \bar{q}$ Bethe-Salpeter bound state is very reasonable for
the distribution function~\cite{Cloet:2005pp,Wakamatsu:1997en}, but the DLY crossing arguments
then imply the truncation of the
spectator state $p_n$, of Eq.~\eqref{ddef}, to a single quark state.
Although this simple assumption does not lead to any violation
of conservation laws, the sum rules in Eqs.~\eqref{mom} and \eqref{iso} cannot
be satisfied in a single step fragmentation process. 

For these reasons, we will not rely on Eq.~\eqref{dly} to calculate
the fragmentation functions, although we will confirm its formal validity 
for the lowest order (elementary) functions. We note that the arguments given 
above do not question the usefulness of Eq.~\eqref{dly} as a means to relate 
the kernels of the $Q^2$ evolution equations for the distribution and
fragmentation functions (see Ref.~\cite{Blumlein:2000wh} and Appendix B). 
In fact, it is known that at leading order (LO)  
in $\alpha_s$ this relation between the kernels is valid, 
although it is violated at next-to-leading order (NLO)~\cite{Stratmann:1996hn}.

\section{Elementary distribution and fragmentation functions}
\label{sec:elementary}

\begin{figure}[tbp]
\centering\includegraphics[width=0.7\columnwidth]{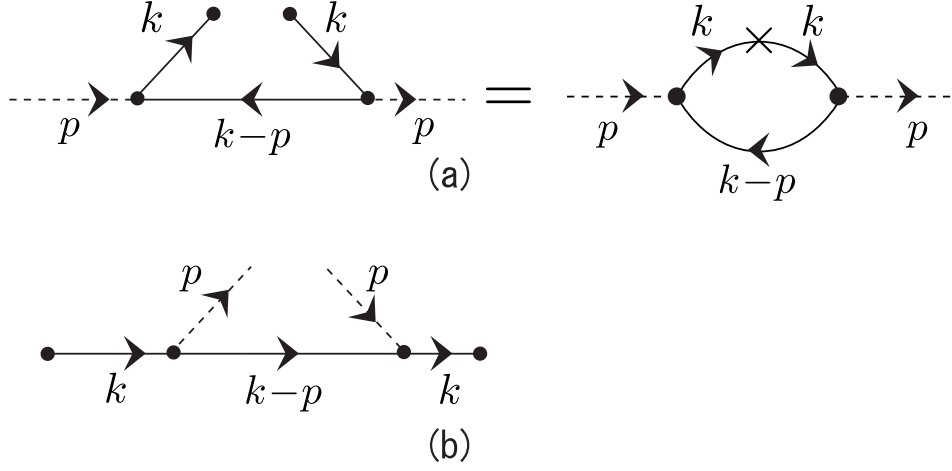}
\caption{Figure (a) depicts the cut diagram (left) and Feynman diagram (right) for the 
distribution function $f_q^{\pi}(x)$. Solid lines denote the quark and
dashed lines the pion. Here $k_- = x\,p_-$ 
and the two quark lines with momentum $k$ are connected by a $\gamma^+$.
Figure (b) depicts the cut diagram for the fragmentation function $d_q^{\pi}(z)$.
Here $p_-= z\,k_-$ and the two quark lines with momentum $k$ are connected by a $\gamma^+$. 
This diagram refers to a frame where ${\vect p}_T=0$ and  
the substitution given in Eq.~\eqref{sub} is performed in the final transverse 
momentum integral.}
\label{fig:feynman_dia_1}
\end{figure}

The elementary distribution and fragmentation functions for the
pion are represented in Figs.~\ref{fig:feynman_dia_1} as cut diagrams. 
Since the distribution function can also be obtained from a straightforward 
Feynman diagram calculation~\cite{Jaffe:1983hp,Bentz:1999gx},\footnote{This 
is seen simply by using completeness 
in Eq.~\eqref{fdef} and the identity $\psi_+(0)^{\dagger} \psi_+(\omega^-)
= T\left(\psi_+(0)^{\dagger} \psi_+(\omega^-)\right)$ in the
limit $\omega^+ \rightarrow 0-\epsilon$, which follows from causality.}
we also illustrate the Feynman diagram for the distribution function 
on the right hand side in Fig.~\ref{fig:feynman_dia_1}a. We denote
the elementary fragmentation function by $d_q^h$ in order to distinguish it 
from the total fragmentation function $D_q^h$ determined in 
Section~\ref{sec:product}.
We obtain the following expressions from the diagrams in 
Figs.~\ref{fig:feynman_dia_1}:\footnote{The expressions given in this 
section refer to the NJL model, however they actually 
have the same form in any effective chiral quark model
with point-like pion-quark vertex functions.}
\begin{align}
\label{fqpi1}
f_q^{\pi}(x) &= \frac{1}{2} \left(1+\tau_{\pi} \tau_q \right) \, 
3 g_{\pi}^2 \int \frac{d^4 k}{(2\pi)^4}\,
\mathrm{Tr}_D \left[ S_F(k) \gamma^+ S_F(k) \gamma_5 \left(\slashed{k}
- \slashed{p} - M \right) \gamma_5 \right] 
\delta(k_--p_- x) \, \delta\!\left((p-k)^2 - M^2 \right), \\
\label{fqpi2} 
&= \frac{1}{2} \left(1+\tau_{\pi} \tau_q \right) \, 
6 g_{\pi}^2 \int \frac{d^2 k_T}{(2\pi)^3}\,
\frac{{\vect k}_T^2 + M^2}{\left[{\vect k}_T^2 + M^2 - m_\pi^2 x(1-x)
\right]^2}, \\
\label{dqpi1}
d_q^{\pi}(z) &= \frac{1}{2} \left(1+\tau_{\pi} \tau_q \right) \, 
g_{\pi}^2 \, \frac{z}{2} \int \frac{d^4 k}{(2\pi)^4}\,
\mathrm{Tr}_D \left[ S_F(k) \gamma^+ S_F(k) \gamma_5 \left(\slashed{k}
- \slashed{p} - M \right) \gamma_5 \right]\,
\delta(k_--p_-/z)\, \delta\!\left((p-k)^2 - M^2 \right), \\
\label{eq:dqpi3}
&\hspace{-2mm}\left[= \frac{z}{6} f_q^{\pi}\left(x = \frac{1}{z} \right)\right], \\
\label{dqpi2}
&= \frac{1}{2} \left(1+\tau_{\pi} \tau_q \right) \,
z \, g_{\pi}^2  \int \frac{d^2 p_{\perp}}{(2\pi)^3}\,
\frac{{\vect p}_{\perp}^2 + M^2 z^2}{\left[{\vect p}_{\perp}^2+M^2 z^2 
+ (1-z)\, m_\pi^2 \right]^2},
\end{align}
where $\mathrm{Tr}_D$ indicates a trace over Dirac indices only.
The Feynman propagator of a constituent quark with mass $M$ is denoted by $S_F$
and $g_{\pi}$ is the pion-quark coupling constant. In the
NJL model $g_{\pi}$ is defined via the residue of the $q \bar{q}$ $t$-matrix
at the pion pole, and can be expressed in terms of the $q \bar{q}$ bubble
graph by
\begin{align}
g_{\pi}^{-2} = -\frac{\partial\, \Pi_{\pi}(q^2)}{\partial q^2}\biggr\rvert_{q^2 = m_\pi^2}, 
\qquad \text{where}
\qquad \Pi_{\pi}(q^2) = 6 i \int \frac{d^4 k}{(2\pi)^4}
\mathrm{Tr}_D \left[\gamma_5 S_F(k) \gamma_5 S_F(k+q) \right].
\label{gpi}
\end{align}
We use the isospin notations $(\tau_u, \tau_d) = (1, -1)$ and
$\left(\tau_{\pi^+}, \tau_{\pi^0},\tau_{\pi^-}\right) = (1, 0, -1)$.
For the distribution function in the physical region ($0<x<1$) 
a factor $\Theta\left(p_--k_-\right) = \Theta(1-x)$ has to be
supplied in Eq.~\eqref{fqpi1}, which expresses the fact that the
intermediate antiquark in Fig.~\ref{fig:feynman_dia_1}a has positive energy. Similarly, for the 
fragmentation function a factor $\Theta(k_- - p_-) = \Theta(1-z)$ 
has to be supplied in Eq.~\eqref{dqpi1}, because the intermediate quark
in Fig.~\ref{fig:feynman_dia_1}b has positive energy. To obtain 
Eq.~\eqref{dqpi2} we made the substitution given in Eq.~\eqref{sub}.

The DLY relation on this level, indicated in brackets as Eq.~\eqref{eq:dqpi3}, 
shows that Eq.~\eqref{fqpi1} can be considered as a
generalized distribution function, which gives the physical
distribution function in the region $0<x<1$ and the fragmentation
function in the region $x= 1/z > 1$. 
The reason why we indicate this relation only in
brackets is that it is violated if the integrals are
regularized. For example, if we use a sharp cut-off ($\Lambda$) 
for the transverse quark momentum in Eq.~\eqref{fqpi2}, a strict application
of the DLY relation would mean that 
the transverse momentum of the produced pion in Eq.~\eqref{dqpi2} should be
cut at $z\,\Lambda$, which is unacceptable. The more physical
procedure is to impose $|{\vect k}_T| < \Lambda$ on Eq.~\eqref{fqpi2} and
$|{\vect p}_{\perp}| < \Lambda$ on Eq.~\eqref{dqpi2}, which breaks the DLY relation.
A similar breakdown of the DLY relation occurs in any other sensible 
regularization scheme. A noticeable consequence of this is that in the
chiral limit the distribution function of Eq.~\eqref{fqpi2} becomes a
constant, but the fragmentation function of Eq.~\eqref{dqpi2} is not
linear in $z$, as the DLY relation indicated in Eq.~\eqref{eq:dqpi3} would suggest.

\begin{figure}[tbp]
\centering\includegraphics[width=0.4\columnwidth]{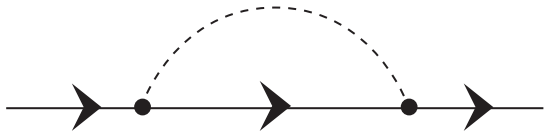}
\caption{The quark self-energy,
$\Sigma_Q^{(\pi)}(k) = -3i {g}_{\pi}^2 
\int \frac{d^4 p}{(2\pi)^4} \gamma_5 S_F(k-p) \gamma_5 \Delta_{F}(p)$, 
where $\Delta_F$ is the Feynman propagator of the pion.}
\label{fig:selfenergy}
\end{figure}

The relations for the distribution function
\begin{align}
\int_0^1 dx \, f_q^{\pi}(x) = \frac{1}{2} \left(1 + \tau_{\pi} \tau_q\right),
\qquad \text{and} \qquad
\int_0^1 dx \, x\,f_q^{\pi}(x) = \frac{1}{2} \left(1 + \tau_{\pi} \tau_q
\right) \cdot \frac{1}{2}, 
\label{sumf}
\end{align}
lead to the usual number and momentum sum rules.
For the elementary fragmentation function the following  
relation is obtained from Eq.~\eqref{dqpi2}:
\begin{align} 
\int_0^1 dz \, d_q^{\pi}(z) = \frac{1}{3} \left(1 + \tau_{\pi} \tau_q\right) 
\left(1-Z_Q\right)  \quad \Longrightarrow \quad
\int_0^1 dz \, \sum_{\tau_{\pi}} \, d_q^{\pi}(z) = 1-Z_Q,
\label{sumd}
\end{align}
where $Z_Q$ is the residue of the quark propagator in the presence of the
pion cloud. It is expressed in terms of the renormalized quark self-energy 
${\Sigma}_Q^{(\pi)}(k)$ of Fig.~\ref{fig:selfenergy} as
\begin{align}
1 - Z_Q &= - \left( \frac{\partial\, 
{\Sigma}_Q^{(\pi)}}{\partial \slashed{k}}\right)_{\slashed{k}=M}
 = - \frac{M}{k_-} \left(\bar{u}_Q(k) \frac{\partial\,
{\Sigma}_Q^{(\pi)}}{\partial k_+} u_Q(k)\right)
= \frac{3}{2} g_{\pi}^2 \int_0^1 z \, dz
\int \frac{d^2 p_{\perp}}{(2\pi)^3}
\frac{{\vect p}_{\perp}^2 + M^2 z^2}{\left[{\vect p}_{\perp}^2+M^2 z^2 
+ (1-z)\, m_\pi^2 \right]^2},   
\label{zq}
\end{align}
where $u_Q$ is the quark spinor ($\bar{u}_Q u_Q = 1$).
Because $Z_Q$ is interpreted as the probability to find a \textit{bare} constituent
quark without the pion cloud, Eq.~\eqref{sumd} 
indicates that the elementary fragmentation function is
normalized to the \textit{number of pions per quark}. This is expected from our
general discussions in Section~\ref{sec:operator} and will be elucidated further below. 
Because typical values of
$Z_Q$ in models based on constituent quarks are between 0.8 and 0.9,
we see from Eq.~\eqref{sumd} that the momentum sum rule 
$\int_0^1 dz \, z\,\sum_{\tau_{\pi}} \, d_q^{\pi}(z)$
will be much smaller than typical empirical values.
For example, the NLO analysis of Ref.~\cite{Hirai:2007cx} found
a momentum sum of $\simeq 0.74$. 
From this we can anticipate that 
the elementary fragmentation functions, $d_q^{\pi}$, 
will be very small compared to the empirical ones (see Section~\ref{sec:results}).

\begin{figure}[tbp]
\centering\includegraphics[width=0.7\columnwidth]{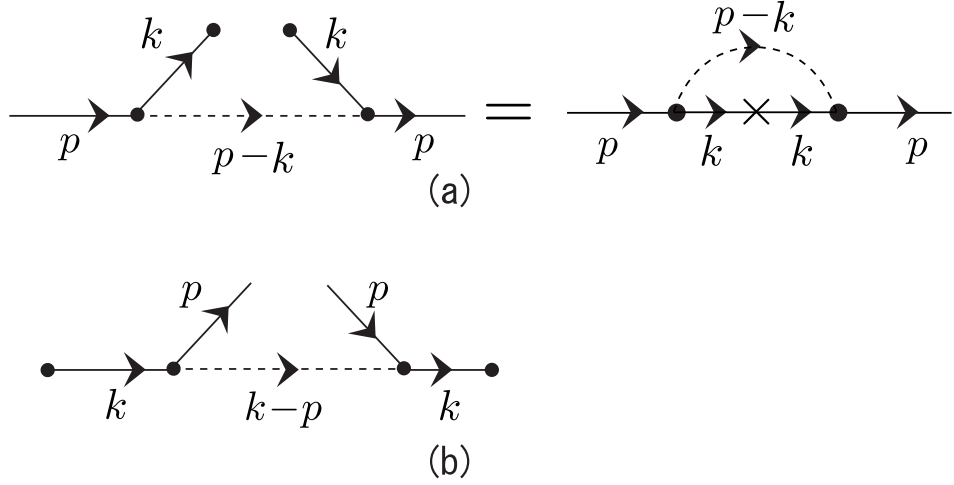}
\caption{Figure (a) depicts the cut diagram (left) and Feynman diagram (right) for the 
loop term in $f_q^{Q}(x)$ of Eq.~\eqref{fqq}. 
Here $k_- = x\,p_-$ 
and the two quark lines with momentum $k$ are connected by a $\gamma^+$. 
Figure (b) depicts the cut diagram for the loop term in $d_q^{Q}(z)$ of Eq.~\eqref{dqq}.
Here $p_- = z\,k_-$ and
the two quark lines with momentum $k$ are connected by a $\gamma^+$. 
This diagram refers to a frame where ${\vect p}_T=0$ and
the substitution in Eq.~\eqref{sub} is performed in the final transverse 
momentum integral.}
\label{fig:feynman_dia_2}
\end{figure}

In order to confirm that this does not mean that momentum conservation
is violated, we also give the expressions for the distribution function of
a quark $q$ inside a parent quark $Q$ and for the fragmentation
function of $q \to Q$. 
The operator definitions of these functions ($f_q^Q(x)$ and $D_q^Q(z)$)
are exactly the same as in Eqs.(1) and (2) with the replacement
$h\to Q$, where the state $\left| p(Q) \right>$ refers to fixed flavour, spin and
color (c.f. the comments in footnote~\ref{foot:3a}). Again we will use the symbol
$d_q^Q$ to denote the elementary fragmentation process.
The relevant cut diagrams are shown in Fig.~\ref{fig:feynman_dia_2}
and a straightforward calculation, following the rules already indicated
in Eqs.~\eqref{fqpi1} and \eqref{dqpi1}, gives\footnote{The tree level terms 
proportional to $Z_Q$ in Eqs.~\eqref{fqq} and
\eqref{dqq} come from the vacuum state in the sum over $n$ in Eqs.~\eqref{fdef} 
and \eqref{ddef}, which contributes for the case where $p(h)$ is a quark.
Using $\psi=\sqrt{Z_Q} \hat{\psi}$, where $\hat{\psi}$ is the 
renormalized quark field with unit pole residue of the propagator, gives the
$Z_Q$ terms in Eqs.~\eqref{fqq} and \eqref{dqq}. Note, in the loop 
terms all factors $Z_Q$ of the propagators cancel. 
We also note that the loop terms in
$f_q^Q(x)$ and $d_q^Q(z)$ formally satisfy the DLY relation, that is
$d_{q,\mathrm{loop}}^Q(z) = (-z/6) f_{q,\mathrm{loop}}^Q(x=1/z)$, however it is
violated after regularization.} 
\begin{align}
\label{fqq}
f_q^Q(x) &= Z_Q \, \delta(x-1)\, \delta_{q,Q}  
+ \left(\frac{1}{2}-\frac{\tau_q \tau_Q}{6} \right) 
\frac{3}{2}\, {g}^2_{\pi} \, (1-x) \int \frac{d^2 k_T}{(2\pi)^3}
\frac{{\vect k}_T^2 + M^2(1-x)^2}{\left[{\vect k}_T^2 + M^2(1-x)^2 
+ x\, m_\pi^2 \right]^2},   \\
\label{dqq} 
d_q^Q(z) &= \frac{1}{6} \, Z_Q \, \delta(z-1) \, \delta_{q,Q} 
+ \frac{1}{6} \left(\frac{1}{2}-\frac{\tau_q \tau_Q}{6} \right)
\frac{3}{2}\, {g}^2_{\pi} \, (1-z) 
\int \frac{d^2 p_{\perp}}{(2\pi)^3}
\frac{{\vect p}_{\perp}^2 + M^2 (1-z)^2}{\left[{\vect p}_{\perp}^2 
+ M^2 (1-z)^2 + z\, m_\pi^2 \right]^2}.  
\end{align}
In accordance with Eq.~\eqref{rel} these relations show that
\begin{align}
d_q^Q(z) = \frac{1}{6} f_q^Q(z) = \frac{1}{6} f_Q^q(z).
\end{align}
Therefore the two quantities 
in Eqs.~\eqref{fqq} and \eqref{dqq} describe essentially the same object,
namely the splitting function of a quark to another quark, which
also includes a ``non-splitting'' term proportional to $Z_Q$. The
normalization is
\begin{align}
 \int_0^1 dz \, 6 \sum_{\tau_Q}\,  d_q^Q(z) = Z_Q + (1-Z_Q) = 1,  
\label{normdqq}
\end{align} 
where the factor 6 represents the summation over the spin and
color of $Q$.
As expected, the second term in Eq.~\eqref{dqq} can be obtained
from the elementary $q \rightarrow \pi$ fragmentation function expressed 
in Eq.~\eqref{dqpi2}, 
via the substitutions $z \rightarrow 1-z$ and 
$\tau_{\pi} \rightarrow (\tau_q - \tau_Q)/2$. 
This directly leads to momentum conservation
for the fragmentation of $q$ into either $Q$ or $\pi$ 
(see Eq.~\eqref{mm2}). 

This connection between splitting functions can also be viewed another way: 
The second term in 
Eq.~\eqref{fqq}, which describes the distribution of $q$ inside $Q$ with a 
pion spectator, suggests that via the substitutions
$\tau_q/2 \rightarrow \tau_Q/2 - \tau_{\pi}$ and $x \rightarrow 1-x$
we obtain the distribution function of a pion inside the quark $Q$, namely
\begin{align}
f_{\pi}^Q (x) =  
 \frac{1}{2} \left(1+\tau_{\pi} \tau_Q \right) \,
g_{\pi}^2 \, x \, \int \frac{d^2 k_{T}}{(2\pi)^3}
\frac{{\vect k}_{T}^2 + M^2 x^2}{\left[{\vect k}_{T}^2+M^2 x^2 
+ (1-x)\, m_\pi^2 \right]^2}. 
\label{fpiq}
\end{align}
Comparison with Eq.~\eqref{dqpi2} gives $d_q^{\pi}(z) = f_{\pi}^q(z)$, 
in accordance with Eq.~\eqref{rel}. This relation further elucidates 
the interpretation of the normalization given in Eq.~\eqref{sumd} as the number of
pions per quark, namely
\begin{align}
 \int_0^1 d z \,\sum_{\tau_{\pi}}\,  d_q^{\pi}(z) =
 \int_0^1 d z \, \sum_{\tau_{\pi}}\, f_{\pi}^{q}(z) = 1-Z_Q.   
\label{inter}
\end{align}

Finally, we write down the momentum sum rules for the elementary
splitting functions. In terms of the distribution functions we have
\begin{align}
\int_0^1 dx\, x \left(\sum_{\tau_q} f_q^Q(x) + \sum_{\tau_{\pi}} f_{\pi}^Q(x) \right)
= Z_Q + \int_0^1 dx\,x \sum_{\tau_{\pi}}
f_{\pi}^Q(1-x) + \int_0^1 dx\, x \sum_{\tau_{\pi}} f_{\pi}^Q(x) = 1,
\label{m1}
\end{align}
where in the second equality we used $x \rightarrow 1-x$ and Eq.~\eqref{inter}.
In terms of the fragmentation functions Eq.~\eqref{m1} becomes
\begin{align}
\int_0^1 dz\, z \left(6 \sum_{\tau_Q} d_q^Q(z) + \sum_{\tau_{\pi}} d_{q}^{\pi}(z) \right) = 1. 
\label{mm2}
\end{align}
In reference to the form of Eq.~\eqref{dqq}, we have the
following simple interpretation of the momentum sum rule of Eq.~\eqref{mm2}:
Because $Z_Q$ is the probability that the initial quark $q$ does not
fragment at all, the fraction $Z_Q$ of the momentum stays with the
initial quark. The remaining fraction $(1-Z_Q)$ is shared among the
quark remainder and the produced pion, that is, the first and second terms 
in Eq.~\eqref{mm2}. 

Although a description of fragmentation functions using 
only the elementary fragmentation processes does not violate any
conservation law, it is completely inadequate for
the following reasons: Firstly, there is a large probability ($Z_Q$)
that the initial quark does not fragment. Secondly, if it does fragment
the momentum fraction $1-Z_Q$ is shared between the quark remainder 
and the pion. Both points are in contradiction to the usual assumption
of complete hadronization, which is expressed by the momentum sum
rule of Eq.~\eqref{mom}.

\section{Generalized Product ansatz for quark cascades}
\label{sec:product}

From the previous section, it is clear that we have to consider the
possibility that the fragmenting quark produces a cascade of mesons.
A simple model to describe cascades is the quark jet-model of Field and
Feynman~\cite{Field:1977fa}. However, the product ansatz used in this model 
assumes from the outset that
the probability for fragmentation in each elementary process is $100\%$,
and that the quark produces an infinite number of mesons. 
Because these
assumptions are inconsistent with our present effective quark theory,
we will first introduce a generalized product ansatz, then explain
its physical significance and its relation to the original quark jet-model.

We assume that the maximum number of mesons which can be produced by
the fragmenting quark is $N$. 
We then consider a process where the initial quark with light-cone momentum
$k_- \equiv W_0$ (which we will simply call the \textit{momentum} in the following) 
goes through a sequence of momenta 
$W_0 \geqslant W_1 \geqslant W_2 \geqslant \dots \geqslant W_N$,
and introduce the momentum ratios  
\begin{align}
\eta_n &= \frac{W_n}{W_{n-1}}, \hspace{10mm} n=1,\dots N .   
\label{eta} 
\end{align}
Our product ansatz for the fragmentation function, which we will motivate
shortly, is:
\begin{multline}
D_q^{\pi}(z) = \sum_{m=1}^{N}  
\int_0^1 d \eta_1 \int_0^1 d \eta_2 \dots \int_0^1 d \eta_N \\
\sum_{Q_N} 6\, d_q^{Q_1}(\eta_1)\cdot 6\, d_{Q_1}^{Q_2}(\eta_2)\cdot \dots \cdot
6\, d_{Q_{N-1}}^{Q_N}(\eta_N)\, \delta\left(z-z_m \right) \, 
\delta\!\left({\tau_{\pi},\left(\tau_{Q_{m-1}}-\tau_{Q_m}\right)/2}\right).
\label{f1}
\end{multline}
Here the functions $d_Q^{Q'}(\eta)$ are our elementary $Q \rightarrow Q'$
splitting functions of Eq.~\eqref{dqq}, which represent the probability
that a quark of flavour $Q$ makes a transition to the quark $Q'$, leaving 
the momentum fraction $\eta$ to $Q'$. A sum over repeated flavour indices
is implied in Eq.~\eqref{f1}; a flavour sum over the quark remainder
($Q_N$) is included; for the case $N=1$ we define $Q_0 \equiv q$; and
the symbol $\delta(i,j)$ denotes the Kronecker delta.
The factor 6 which multiplies each elementary
splitting function comes from the sum over spin and color. The
delta function in Eq.~\eqref{f1} selects a meson, which is produced in the
$m^{\text{th}}$ step with momentum fraction $z_m$ of the initial quark:
\begin{align}
\label{zm1}
z_m = \frac{W_{m-1} - W_m}{W_0} = \eta_1 \cdot \eta_2 \cdot \dots 
\cdot \eta_{m-1}\cdot (1-\eta_m), \quad \text{where} \quad m>1, 
\qquad \text{and} \qquad 
z_1 =  1-\eta_1.  
\end{align}
Because the pion has a mass we will exclude the unphysical case of $z=0$, 
that is, whenever a pion is produced
in the $m^{\text{th}}$ step we will assume that $\eta_m \neq 1$ in Eq.~\eqref{zm1}.

We will write the $q \rightarrow Q$ splitting function of Eq.~\eqref{dqq},
including the spin-color factor 6, in the form 
\begin{align}
\label{two}
6 \, d_q^{Q}(z) &= Z_Q \, \delta(z-1) \, \delta_{q,Q} + F_q^{Q}(z),  
%
%
\end{align}
where 
\begin{align}
F_q^{Q}(z) = \left(\frac{1}{2} - \frac{\tau_q \tau_{Q}}{6} \right) F(z),  
\qquad \text{and} \qquad
F(z) &= \frac{3}{2}\, {g}_{\pi}^2 \, (1-z) \int \frac{d^2 p_{\perp}}{(2\pi)^3}
\frac{\vect{p}_{\perp}^2 + M^2 (1-z)^2}{\left[\vect{p}_{\perp}^2 
+ M^2 (1-z)^2 + z\,m_\pi^2 \right]^2}.   
\label{F2}
\end{align}
The function $F$ satisfies the normalization (see Eq.~\eqref{normdqq})
\begin{align}
\sum_Q \int_0^1 dz\, F_q^Q(z) = \int_0^1 d z \, F(z) = 1-Z_Q.     
\label{Fnorm}
\end{align}
For the case $N=1$ it is easy to see that Eq.~\eqref{f1} reduces
to the elementary fragmentation function of Eq.~\eqref{dqpi2}, namely
\begin{align}
D_q^{\pi}(z) \stackrel{N=1} \longrightarrow F_q^Q(1-z) 
|_{\tau_Q=\tau_q-2\tau_{\pi}} = \frac{1}{3} \left(1 + \tau_q \tau_{\pi}\right)
F(1-z) = d_q^{\pi}(z).
\label{N1}
\end{align}

\begin{figure}[tbp]
\centering\includegraphics[width=0.8\columnwidth]{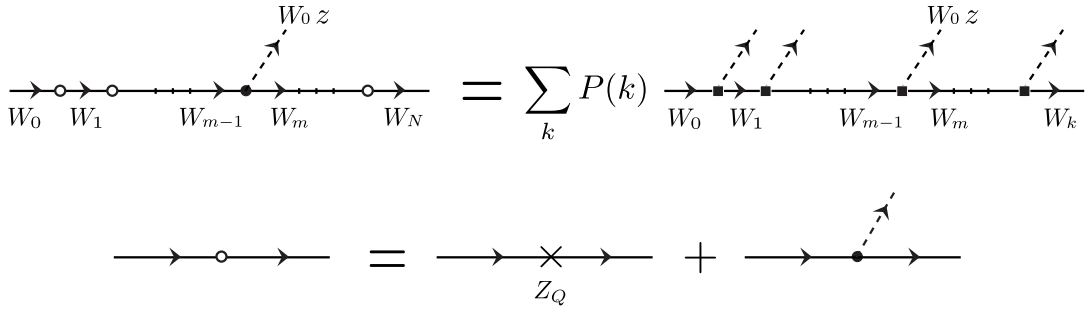}
\caption{The left hand side of the top figure is a graphical representation of
Eq.~\eqref{f1} and the right hand side of this figure represents Eq.~\eqref{y}.
The open circles denote the elementary $q \rightarrow Q$ fragmentation 
function of Eq.~\eqref{two} and the dots represent the second 
(meson emission) term in Eq.~\eqref{two}.
In the $m^{\text{th}}$ step, where a meson with momentum $z\,W_0$ is selected by
the delta-function in Eq.~\eqref{f1}, only the meson emission term contributes. 
The term $P(k)$ is the
binomial distribution of Eq.~\eqref{prob} and the squares represent the 
renormalized meson emission term, $\hat{F}_q^Q(z)$, given by Eq.~\eqref{Fhat}.
The bottom figure is a graphical representaion of Eq.~\eqref{two}.}
\label{fig:productansatz}
\end{figure}

In order to illustrate the physical content of the ansatz expressed by Eq.~\eqref{f1} we 
rewrite it identically as follows:
Noting that each factor of the product in Eq.~\eqref{f1} consists of the two
terms in Eq.~\eqref{two}, it is easy to see that all products with the same
number (call it $k$) of $F's$ and $(N-k)$ number of $Z_Q$'s make the
same contribution to $D_q^{\pi}(z)$. Therefore we can introduce
an ordering of the $\eta$'s in Eq.~\eqref{f1}. That is, take the first $k$
$\eta$'s not equal to one ($\eta_1, \eta_2, \dots \eta_k \neq 1$), 
and the remaining $\eta$'s equal to one ($\eta_{k+1}, \eta_{k+2}, \dots \eta_N = 1$), 
multiply by the combinatoric factor $\binom{N}{k}$ and perform a sum over $k$. 
For some fixed $k$, only terms with $m \leqslant k$ will contribute to the
sum in Eq.~\eqref{f1}, because $z_m$ of Eq.~\eqref{zm1} must be non-zero.\footnote{ 
As explained earlier, we only consider the case $z>0$.} 
Then Eq.~\eqref{f1} is rewritten identically as 
\begin{align}
D_q^{\pi}(z) &= \sum_{m=1}^N \sum_{k=m}^N \, P(k) \, \int_0^1 d \eta_1
\int_0^1 d \eta_2 \dots \int_0^1 d \eta_k  \nonumber \\
&\hspace{60mm}\sum_{Q_k}\, \hat{F}_q^{Q_1}(\eta_1)\, \hat{F}_{Q_1}^{Q_2}(\eta_2) \dots 
\hat{F}_{Q_{k-1}}^{Q_k} (\eta_k) 
\,\delta(z-z_m)\, \delta\! \left(\tau_{\pi},\left(\tau_{Q_{m-1}}-\tau_{Q_m}\right)/2 \right), \nonumber \\
&\equiv \sum_{m=1}^N D_{q,(m)}^{\pi}(z), 
\label{y}
\end{align}
which is expressed graphically in Fig~\ref{fig:productansatz}. The binomial distribution
\begin{align}
P(k) = \binom{N}{k} Z_Q^{N-k} (1-Z_Q)^k,
\label{prob}  
\end{align}
is the probability of producing $k$ mesons out of a maximum of $N$ mesons and
satisfies the normalization condition 
\begin{align}
\sum_{k=0}^N P(k) = 1.  
\label{norm}
\end{align}
In Eq.~\eqref{y} we defined the renormalized function 
$\hat{F}_q^{Q} \equiv F_q^{Q}/(1-Z_Q)$, that is  (see Eqs.~\eqref{F2} and \eqref{Fnorm})
\begin{align}
\hat{F}_q^{Q}(z) &= 
\left(\frac{1}{2} - \frac{\tau_q \tau_{Q}}{6} \right) \hat{F}(z), \qquad \text{where} \qquad
\hat{F}(z) = \frac{F(z)}{1-Z_Q},  \qquad \text{and} \label{Fhat} \\ 
\int_0^1 d z \, \sum_Q \, \hat{F}_q^Q(z) &= \int_0^1 d z \, 
\hat{F}(z) = 1.     
\label{Fhatnorm}
\end{align}

The physical interpretation of Eq.~\eqref{y} is as follows:
\begin{itemize}
\item $P(k)$ is the probability that $k$ mesons out of a maximum of $N$ mesons are produced.
\item $\hat{F}_Q^{Q'}(\eta)$ is the probability density that, \textit{if} a meson 
is emitted from the quark $Q$, the momentum fraction $\eta$ is left to the remaining quark $Q'$.
\item The product $\hat{F}(\eta_1) \cdot \hat{F}(\eta_2) \dots \hat{F}(\eta_k)$ 
is the probability density that, \textit{if} $k$
mesons are produced, each meson carries its momentum fraction $z_m$
($m=1,\dots k$) of the original quark, where $z_m$ is given by Eq.~\eqref{zm1}.
\item $D_{q,(m)}^{\pi}(z)$ is the probability density that the 
$m^\text{th}$ meson has the momentum fraction $z$ of the original quark. 
This implies that at least $m$ mesons must be produced, which corresponds
to the lower limit ($k=m$) of the summation in Eq.~\eqref{y}.
The total fragmentation function $D_q^{\pi}(z)$ is then obtained by summing 
the probability densities $D_{q,(m)}^{\pi}(z)$.   
\end{itemize}
We note that the original ansatz of Field and Feynman~\cite{Field:1977fa} is an
\textit{infinite} product, which formally   
emerges from Eq.~\eqref{y} if we take the limit $N \rightarrow \infty$ and
assume that $P(k)$ is equal to zero for any finite $k$, that is, the
probability of the fragmenting quark to emit a finite number of mesons 
is zero.  
 
We now proceed with Eq.~\eqref{y} in order to find the integral equation
satisfied by the fragmentation function. For a fixed $m$, we can
integrate over $\eta_{m+1}, \dots \eta_N$ by using the
normalization of $\hat{F}$, that is,
\begin{align}
\int_0^1 d\eta \, \sum_Q \, \hat{F}_q^Q(\eta) = 
\int_0^1 d\eta \int_0^1 d\eta' \, \sum_{Q'} \,\hat{F}_q^Q(\eta)
\hat{F}_Q^{Q'}(\eta') = \dots =1.  \label{n}
\end{align}
Then for all $k \geqslant m$ the integrations over the same variables 
$\eta_1, \dots \eta_m$
remain, and the sum over $k$ refers only to the probabilities $P(k)$.
Performing the shift $\eta_m \rightarrow 1 - \eta_m$ in the integral
over $\eta_m$, we obtain
\begin{multline}
D_{q(m)}^{\pi}(z) = \left( \sum_{k=m}^N P(k) \right) \int_0^1 d\eta_1
\int_0^1 d\eta_2 \dots \int_0^1 d\eta_m \\
\hat{F}_q^{Q_1}(\eta_1) \hat{F}_{Q_1}^{Q_2}(\eta_2) \dots 
\hat{F}_{Q_{m-2}}^{Q_{m-1}}\left(\eta_{m-1}\right)\,
\hat{d}_{Q_{m-1}}^{\pi}(\eta_m) \, \delta(z-\eta_1 \eta_2 \dots \eta_m).   
\label{y1}
\end{multline}
The function $\hat{d}_q^{\pi}(z) \equiv d_q^{\pi}(z)/(1-Z_Q)$ is
the renormalized elementary $q \rightarrow \pi$ fragmentation function, 
therefore (see Eq.~\eqref{N1})
\begin{align}
\hat{d}_q^{\pi}(z) = \hat{F}_q^{Q}(1-z) |_{\tau_Q = \tau_q - 2 \tau_{\pi}} 
 = \frac{1}{3} \left( 1 + \tau_q \tau_{\pi}\right) \hat{F}(1-z).   
\label{relat} 
\end{align}
From Eq.~\eqref{y1} it is easy to derive the following recursion relation
for $m>1$:
\begin{align}
D_{q(m)}^{\pi}(z) = R_m \left[ \hat{F}_q^Q  \otimes D_{Q(m-1)}^{\pi}\right](z),
\qquad \text{where} \qquad m>1,
\label{dm}
\end{align}
while for $m=1$ we have
\begin{align}
D_{q(1)}^{\pi}(z) = R_1 \, \hat{d}_q^{\pi}(z).
\label{d1}
\end{align}
We have introduced the following ratios:
\begin{align}
R_n = \frac{\sum_{k=n}^N P(k)}{\sum_{k=n-1}^N P(k)}, 
\qquad \text{where} \qquad
n=1,2,\dots N,   
\label{rr}
\end{align}
and used the following notation for the convolution of two functions $A(z)$ and $B(z)$:
\begin{align}
\left[ A \otimes B \right](z) = \int_0^1 dz_1 \int_0^1 dz_2
 \, \delta(z - z_1 z_2) A(z_1) B(z_2).  \label{conv}
\end{align}
The total fragmentation function then becomes
\begin{align}
D_q^{\pi}(z) = R_1 \, \hat{d}_q^{\pi}(z) + \sum_{n=2}^N \, R_n \left[
\hat{F}_{q}^Q \otimes  D_{Q (n-1)}^{\pi} \right](z) ,   \label{fin1}
\end{align}
where $D_{q(m)}^{\pi}$ can be obtained from the recursion relation 
of Eq.~\eqref{dm}, with the starting value given by Eq.~\eqref{d1}.

It is interesting at this stage to derive the sum rules for the 
fragmentation function. A simple calculation using Eq.~\eqref{y1} 
gives the following expressions for the multiplicity, the momentum sum 
and the isospin sum:
\begin{align}
\label{mult}
\int_0^1 d z \, \sum_{\tau_{\pi}} \, D_q^{\pi}(z) &= 
\sum_{k=1}^N k P(k) =  N (1-Z_Q),  \\
\label{ms} 
\int_0^1 d z \, \sum_{\tau_{\pi}} z\, D_q^{\pi}(z) &=
1 - \sum_{k=0}^N P(k) \langle z \hat{F} \rangle^k 
= 1 - \left( Z_Q + (1-Z_Q) \langle z \hat{F} \rangle \right)^N , \\
\label{is}
\int_0^1 d z \, \sum_{\tau_{\pi}} \, \tau_{\pi} \, D_q^{\pi}(z) &= 
\frac{\tau_q}{2} \left[ 1 - \sum_{k=0}^N P(k) \left(- \frac{1}{3}\right)^k \right]
= \frac{\tau_q}{2} \left[ 1 - \left( Z_Q - \frac{1}{3} (1-Z_Q) \right)^N\right],  
\end{align}
where $\langle A \rangle \equiv \int_0^1 dz A(z)$.
These expressions can be understood as follows: 
If $k$ mesons are produced with probability $P(k)$, then Eq.~\eqref{mult}
is simply the mean number of mesons; the quantity  
$P(k) \langle z \hat{F} \rangle^k$ in Eq.~\eqref{ms} is the 
mean momentum fraction left to the quark remainder; and the quantity 
$P(k) \left(- 1/3 \right)^k$ in Eq.~\eqref{is} is the
mean isospin fraction left to the quark remainder.

Eqs.~\eqref{ms} and \eqref{is} indicate that, in the present model,
it is not possible to transfer the total momentum and isospin of the
original quark to the mesons, if the maximum number of mesons is finite.
The momentum and isospin sum rules given in Eqs.~\eqref{mom} and \eqref{iso} are
valid only in the limit $N\rightarrow \infty$. While this may indicate
a conceptual limitation of the jet-model, we note that in general, the
QCD based empirical analysis of fragmentation functions also leads to divergent
multiplicities. Therefore, we find it more important to satisfy the
momentum and isospin sum rules given in Eqs.~\eqref{mom} and \eqref{iso} than to
have finite multiplicities, and therefore we take the limit $N\rightarrow \infty$. 
The results then become
independent of the form of the distribution $P(k)$, if the following 
condition is satisfied for the ratios in Eq.~\eqref{rr}:
\begin{align}
R_n \stackrel{N\rightarrow \infty}{\longrightarrow} 1, 
\qquad \text{for all} \quad n=1,2,\dots
\label{cond}
\end{align}
In fact, it is well known that in the 
limit $N\rightarrow \infty$ our binomial 
distribution of Eq.~\eqref{prob} becomes a normalized Gaussian distribution
(normal distribution) 
$\frac{1}{\sqrt{2\pi c^2}} \, e^{- \frac{(k-k_0)^2}{2c^2}}$, 
with the same mean value $k_0=N(1-Z_Q)$ and variance $c^2=NZ_Q(1-Z_Q)$
as the original binomial distribution.
The validity of Eq.~\eqref{cond} can then easily be confirmed. In fact, 
\textit{any} distribution which approaches a normal distribution
in the limit $N\rightarrow \infty$ satisfies the 
condition given in Eq.~\eqref{cond}.\footnote{The fact that in the limit $N\rightarrow \infty$
the binomial distribution becomes a normal distribution is known as the Moivre-Laplace 
theorem, which can be formulated rigorously in integral form
(``weak convergence''). The central limit
theorem~\cite{CL} is an extension of the Moivre-Laplace theorem to general
distributions $P(k)$ with mean value proportional to $N$ and variance $c^2 \propto
N$. This indicates that Eq.~\eqref{cond} is actually valid for a wide class
of distributions. Although our NJL-jet model ansatz of Eq.~\eqref{f1}
leads to the binomial distribution, in the limit $N \rightarrow \infty$
the results hold for a wide class of distributions.}

Using Eq.~\eqref{cond}, we see from Eq.~\eqref{fin1} that our fragmentation 
function satisfies essentially the same integral equation as in
the original quark jet-model~\cite{Field:1977fa}:
\begin{align}
D_q^{\pi}(z) = \hat{d}_q^{\pi}(z) + \left[\hat{F}_q^Q \otimes D_Q^{\pi}\right](z),
\label{final}
\end{align}
where the driving term is given by Eq.~\eqref{relat} and the integral
kernel by Eq.~\eqref{Fhat}.
We finally write down the equations which we solve in the next section.
Defining two functions $A(z)$ and $B(z)$ by the isospin decomposition 
\begin{align}
D_q^{\pi}(z) \equiv \frac{1}{3} \left[A(z) + \tau_q \tau_{\pi} B(z) \right],
\label{ab}
\end{align}
and using Eqs.~\eqref{Fhat} and \eqref{relat}, we find the following integral
equations for $A(z)$ and $B(z)$ from Eq.~\eqref{final}: 
\begin{align}
A(z) &= \hat{F}(1-z) + \int_z^1 \frac{dy}{y} \, \hat{F}
\left(\frac{z}{y}\right) \, A(y), \label{a} \\
B(z) &= \hat{F}(1-z) - \frac{1}{3} \int_z^1 \frac{dy}{y} \, \hat{F}
\left(\frac{z}{y}\right) \, B(y), \label{b}
\end{align}
where $\hat{F}(z)$ is obtained by renormalizing the function $F(z)$
in Eq.~\eqref{F2} to unity.
Using Eq.~\eqref{ab}, we obtain the following expressions for the 
\textit{favoured}, \textit{unfavoured} and \textit{neutral} fragmentation functions:
\begin{align}
D_u^{\pi^+} &= D_d^{\pi^-} = D_{\bar{u}}^{\pi^-} = 
D_{\bar{d}}^{\pi^+} = \frac{1}{3}\left(A+B\right) ,  \label{fav} \\  
D_u^{\pi^-} &= D_d^{\pi^+} = D_{\bar{u}}^{\pi^+} = 
D_{\bar{d}}^{\pi^-} = \frac{1}{3}\left(A-B\right),  \label{unfav} \\  
D_u^{\pi^0} &= D_d^{\pi^0} = D_{\bar{u}}^{\pi^0} = 
D_{\bar{d}}^{\pi^0} = \frac{1}{3} \, A .              \label{neut}  
\end{align}
%
From the form of Eqs.~\eqref{a} and \eqref{b} it is
easily seen that $\langle z\,A \rangle =1$ and $\langle B \rangle = 3/4$, 
which leads to the
momentum and isospin sum rules of Eqs.~\eqref{mom} and \eqref{iso}. For
large $z$, both functions $A(z)$ and $B(z)$ approach $\hat{F}(1-z)$ and
therefore the unfavored fragmentation functions
in Eq.~\eqref{unfav} are suppressed for large pion momenta.

\section{Numerical results and discussions}
\label{sec:results}

In this section we present the numerical results for the fragmentation
function of Eq.~\eqref{final} in the NJL-jet model. For reference, we also
give the results for the elementary distribution function of Eq.~\eqref{fqpi2}. 
Because the application of the NJL model to the calculation of the quark distribution
functions in the pion has been explained in detail in Ref.~\cite{Bentz:1999gx}, we will not 
repeat the explanations of the model here. For convenience,
we will use the same regularization scheme, namely the invariant mass,
or Lepage-Brodsky (LB)~\cite{Lepage:1980fj} regularization scheme, 
with the same parameters as in Ref.~\cite{Bentz:1999gx}. 
The LB scheme is suitable for regularizing integrals
in terms of light-cone variables and in terms of the usual variables it is
equivalent to the familiar 3-momentum cut-off scheme~\cite{Bentz:1999gx}. That is, if we
denote the 3-momentum cut-off by $\Lambda_3$, which is fixed in the usual
way by reproducing the experimental pion decay constant, a bubble-type 
loop integral with two intermediate particles of mass $M_1$ and $M_2$ 
is regularized by cutting off their invariant mass $M_{12}$ according to
\begin{align}
M_{12} \leqslant \Lambda_{12} \equiv \sqrt{\Lambda_3^2 + M_1^2} + 
\sqrt{\Lambda_3^2 + M_2^2}.   \label{inv}
\end{align}
In terms of light-cone variables, if we associate with particle 1 the
transverse momentum $\vect{q}_T$ and the momentum 
fraction $y$ of the total $P_-$ momentum, 
and to particle 2 we associate the momentum  fraction $(1-y)$ and 
transverse momentum $-\vect{q}_T$, then their invariant mass squared is
\begin{align}
M_{12}^2 = \frac{M_1^2 + \vect{q}_T^2}{y} + \frac{M_2^2 + \vect{q}_T^2}{1-y}.
\label{inv1}
\end{align}
The requirement $M_{12} \leqslant \Lambda_{12}$ then leads to a $y$-dependent transverse
cut-off: $\vect{q}_T^2 \leqslant \Lambda_{12}^2 \, y(1-y) - M_1^2(1-y) - M_2^2 y$.
This condition also restricts the values of $y$ from below and above
($0<y_1 \leqslant y \leqslant y_2 <1$). 
For example, for the integral in Eq.~\eqref{dqpi2} of the
elementary $q \rightarrow \pi$ fragmentation function we have $M_1=m_\pi$ and $M_2=M$,
for the integral in Eq.~\eqref{dqq} of the elementary $q \rightarrow Q$ fragmentation
function we have $M_1 = M$ and $M_2=m_\pi$ and for the integral in Eq.~\eqref{fqpi2} of
the distribution function we have $M_1=M_2=M$. We also note that this
regularization scheme does not violate the sum rules.  

Following Ref.~\cite{Bentz:1999gx} we use a constituent quark mass of $M=300$ MeV. 
Then $\Lambda_3 = 670$ MeV and the invariant mass cut-offs for the 
$(\pi,q)$ and $(q,q)$ systems are 1.42 GeV and 1.47 GeV, respectively. 
We did not investigate whether other parameter sets or other regularization
schemes lead to a better description of the fragmentation functions. 

As usual, we will associate a low energy renormalization scale ($Q_0^2$)
to our NJL results and evolve them in $Q^2$ by using
the QCD evolution equations. For the evolution of the 
fragmentation functions we limit ourselves to LO. 
In this case it has been verified~\cite{Blumlein:2000wh} that
a formal application of the DLY relation, see Eq.~\eqref{dly}, leads to the
correct connection between the evolution kernels of the distribution and
fragmentation functions (see Appendix B). However, the DLY relation is 
not actually used to relate the distribution and fragmentation functions 
themselves. We therefore use the $Q^2$
evolution code of Ref.~\cite{Miyama:1995bd} at LO for the distribution functions, and 
perform the transformation of the kernels as explained in Appendix B 
to obtain the LO evolution of the fragmentation functions.\footnote{The DLY 
based relation between the evolution kernels for distribution and
fragmentation functions is violated at NLO~\cite{Blumlein:2000wh}. 
Unfortunately, a NLO evolution code
for the fragmentation functions is not yet publicly available.
In this paper we do not attempt a quantitative comparison with the 
empirical functions, therefore we leave the NLO calculation for future work.}
 
\begin{figure}[tbp]
\subfigure{\includegraphics[width=0.48\columnwidth,clip=true,angle=0]{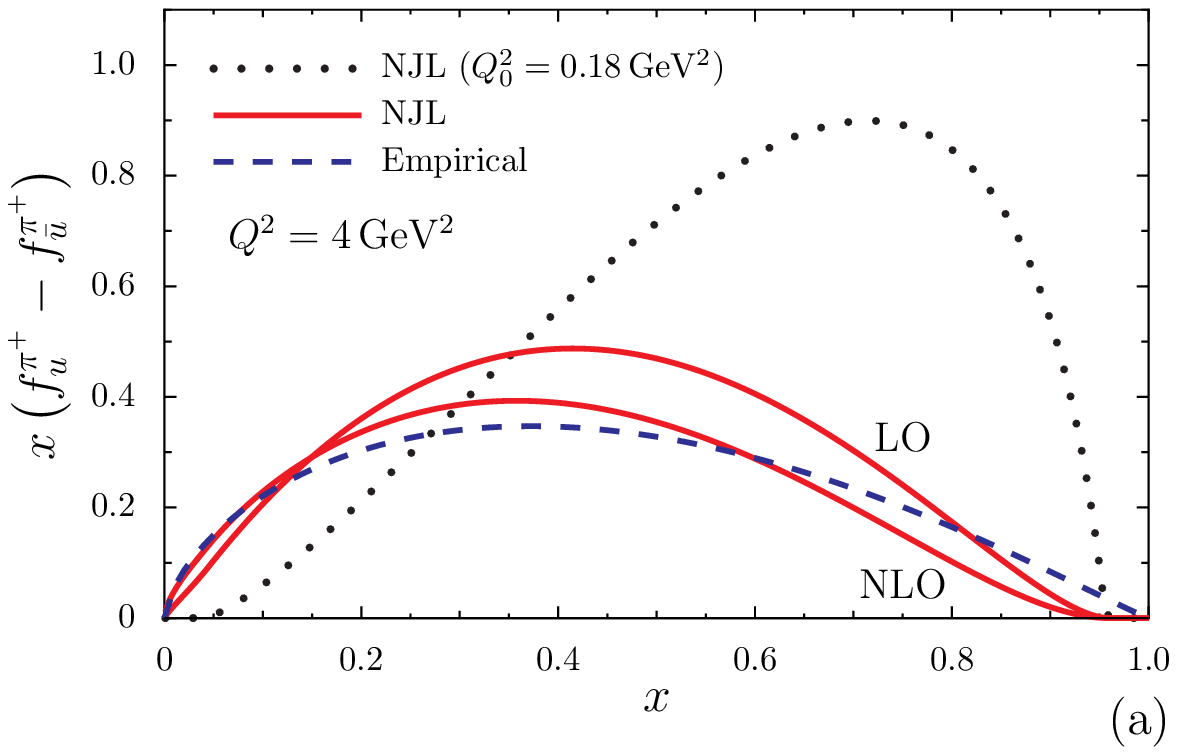}} \quad
\subfigure{\includegraphics[width=0.48\columnwidth,clip=true,angle=0]{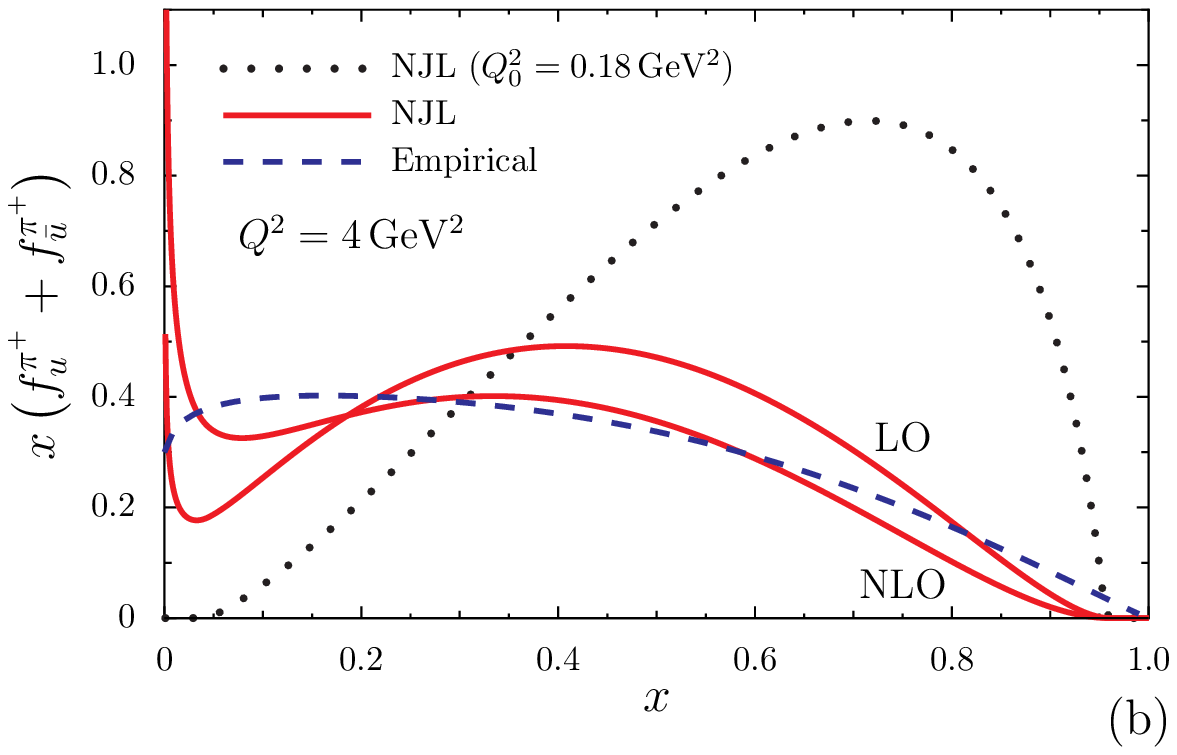}} 
\vskip -1em
\caption{
Figure (a) depicts the \textit{minus-type} (valence) quark 
distribution $x(f_u^{\pi^+}(x)-f_{\bar{u}}^{\pi^+}(x))$ and
figure (b) illustrates the \textit{plus-type} quark distribution 
$x(f_u^{\pi^+}(x)+f_{\bar{u}}^{\pi^+}(x))$ of the $u$-quark in a $\pi^+$. 
The dotted line is the NJL model result, used as input ($Q_0^2=0.18$ GeV$^2$) for the 
$Q^2$ evolution. The solid line labeled by LO (NLO) is the result of LO (NLO)
evolution to $Q^2=4$ GeV$^2$. 
The dashed line is the empirical NLO result of Ref.~\cite{Sutton:1991ay}
at $Q^2=4$ GeV$^2$.}  
\label{fig:qdis}
\end{figure}

In Fig.~\ref{fig:qdis}a we recapitulate the results of Fig.~4 of Ref.~\cite{Bentz:1999gx}, 
and show the \textit{minus-type} (valence, $q-\bar{q}$) $u$-quark distribution
in a $\pi^+$ and in Fig.~\ref{fig:qdis}b we give the result for the 
\textit{plus-type} ($q+\bar{q}$) $u$-quark distribution in a $\pi^+$.  
The dotted line shows the NJL model result based on Eq.~\eqref{fqpi2}, the solid 
lines illustrate the distribution obtained by associating
a low energy scale of $Q_0^2=0.18$ GeV$^2$ to the NJL result and
performing the $Q^2$ evolution at LO and NLO to $Q^2 = 4$ GeV$^2$. 
The dashed line shows the empirical NLO parametrizations of 
Ref.~\cite{Sutton:1991ay}. We see that the LO and NLO results show 
quantitative differences because of the rather low value assumed for $Q_0^2$,
although the qualitative behaviours are similar. 

\begin{figure}[tbp]
\subfigure{\includegraphics[width=0.48\columnwidth,clip=true,angle=0]{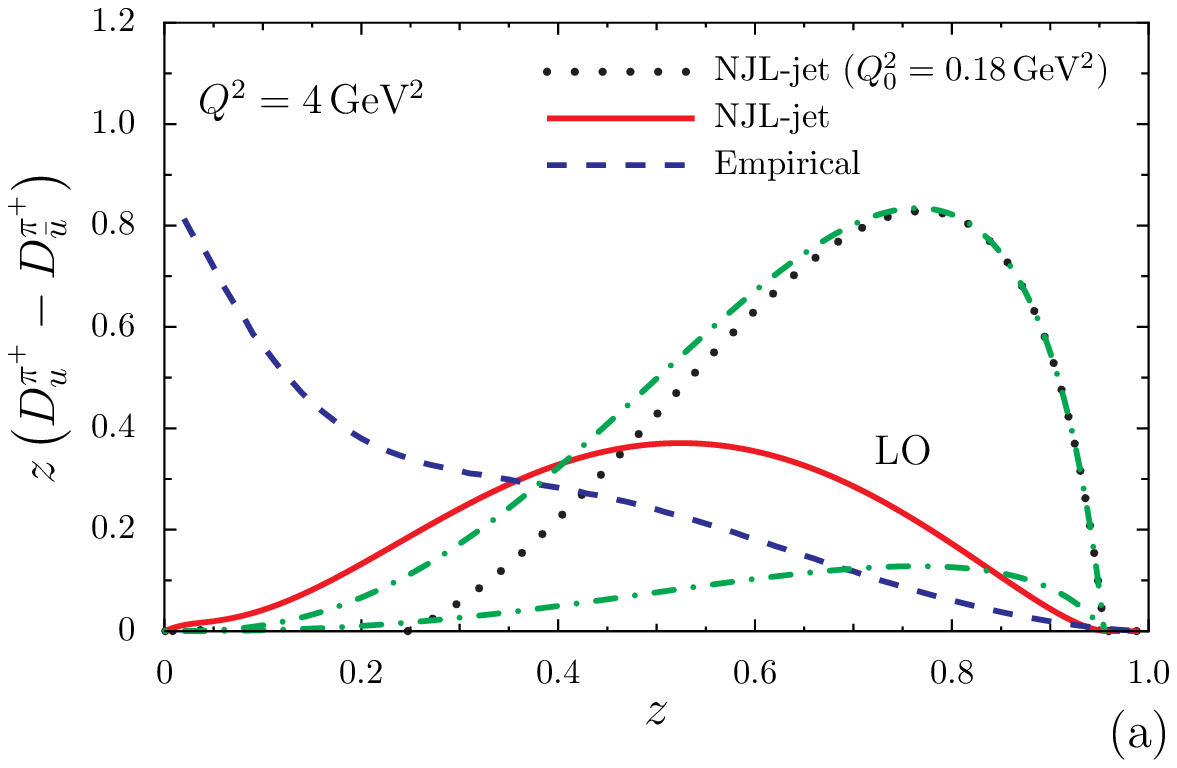}} \quad
\subfigure{\includegraphics[width=0.48\columnwidth,clip=true,angle=0]{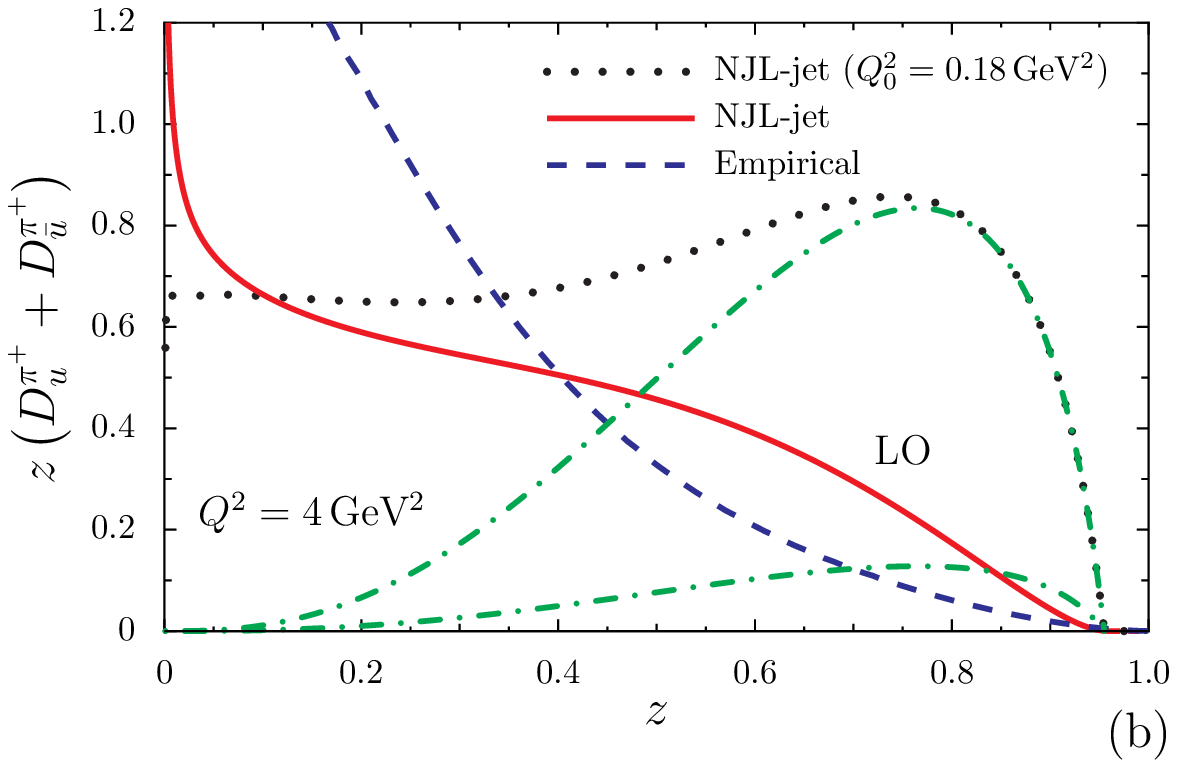}} 
\vskip -1em
\caption{Figure (a) depicts the \textit{minus-type} fragmentation function
$z(D_u^{\pi^+}(z)-D_{\bar{u}}^{\pi^+}(z))$ and
figure (b) illustrates the \textit{plus-type} fragmentation function 
$z(D_u^{\pi^+}(z)+D_{\bar{u}}^{\pi^+}(z))$ for $u \rightarrow \pi^+$. 
The dotted line is the NJL-jet model result, used as input ($Q_0^2=0.18$ GeV$^2$) for the 
$Q^2$ evolution. The lower dash-dotted line is the elementary 
fragmentation function ($d^\pi_q$ of Eq.~\eqref{dqpi2}) and the upper dash-dotted
line is the renormalized elementary fragmentation function ($\hat{d}^\pi_q$ of Eq.~\eqref{relat}), 
which is the driving term of the integral equation expressed in Eq.~\eqref{final}. 
The solid line is the result after LO evolution to $Q^2=4$ GeV$^2$. The
dashed line is the empirical NLO result of Ref.~\cite{Hirai:2007cx}, evolved to $Q^2=4$ GeV$^2$.}  
\label{fig:frag}
\end{figure}

In Figs.~\ref{fig:frag} we present the corresponding results for the minus-type and
plus-type fragmentation functions for $u \rightarrow \pi^+$. The NJL-jet 
result, given by the dotted line, is the solution of the integral equation
in Eq.~\eqref{final}. Therefore the dotted line in Figs.~\ref{fig:frag}a 
and \ref{fig:frag}b show the functions $\frac{2}{3} B(z)$ and $\frac{2}{3} A(z)$,
respectively (see Eqs.\eqref{fav} and \eqref{unfav}).
In order to see the importance of the cascade processes, we
also plot the driving term of the integral equation, namely
$\frac{2}{3} \hat{F}(1-z)$, as the upper dash-dotted line, which
is the renormalized elementary fragmentation function. As the 
lower dash-dotted line we illustrate the elementary fragmentation function,
namely $\frac{2}{3} F(1-z)$.
The result of the evolution of the dotted line ($Q_0^2=0.18$ GeV$^2$) to 
$Q^2 = 4$ GeV$^2$ at LO is shown by the solid line and the dashed line shows the 
empirical NLO result of Ref.~\cite{Hirai:2007cx}, evolved to $Q^2=4$ GeV$^2$.

Several important points are illustrated in Figs.~\ref{fig:frag}. Firstly, as
anticipated in Section~\ref{sec:elementary}, the elementary fragmentation function 
(lower dash-dotted line) is very
small. Secondly, Fig.~\ref{fig:frag}b shows the tremendous
enhancement at intermediate and small $z$ of the plus-type fragmentation
function caused by the cascade processes (iterations of the integral equation 
of Eq.~\eqref{final}),
while for the minus-type fragmentation function of Fig.~\ref{fig:frag}a a small reduction is
seen. Thirdly, the calculated result shown by the solid line has the
correct order of magnitude for intermediate and large $z$, when compared
with the empirical function. This point, which reflects the fact that our
model satisfies the momentum sum rule, is very important, because
effective quark model calculations completed hitherto only considered the 
elementary fragmentation functions and introduced some ad hoc parameters 
(like normalization constants) to obtain the correct order of
magnitude. Quantitatively, Figs.~\ref{fig:frag} indicate that our fragmentation
functions are too big at large $z$ and too small at smaller $z$. 
This is natural for the following reasons: Firstly, we can expect that a NLO 
calculation will lead to a softening of the fragmentation functions. Secondly, some of 
the observed pions are secondary ones, which come from the decay of primary
$\rho$ and $\omega$ mesons. Thirdly, the coupling to other fragmentation channels, 
in particular the
nucleon, antinucleon and kaon, will transfer some amount of the hard quark momentum 
to these other hadrons.
Also, one should not forget that the empirical fragmentation functions
have very large uncertainties, which are not indicated in our
figures. Nevertheless, Figs.~\ref{fig:frag} indicate that the present
NJL-jet model provides a reasonable starting point for the
description of fragmentation functions.

\begin{figure}[tbp]
\subfigure{\includegraphics[width=0.48\columnwidth,clip=true,angle=0]{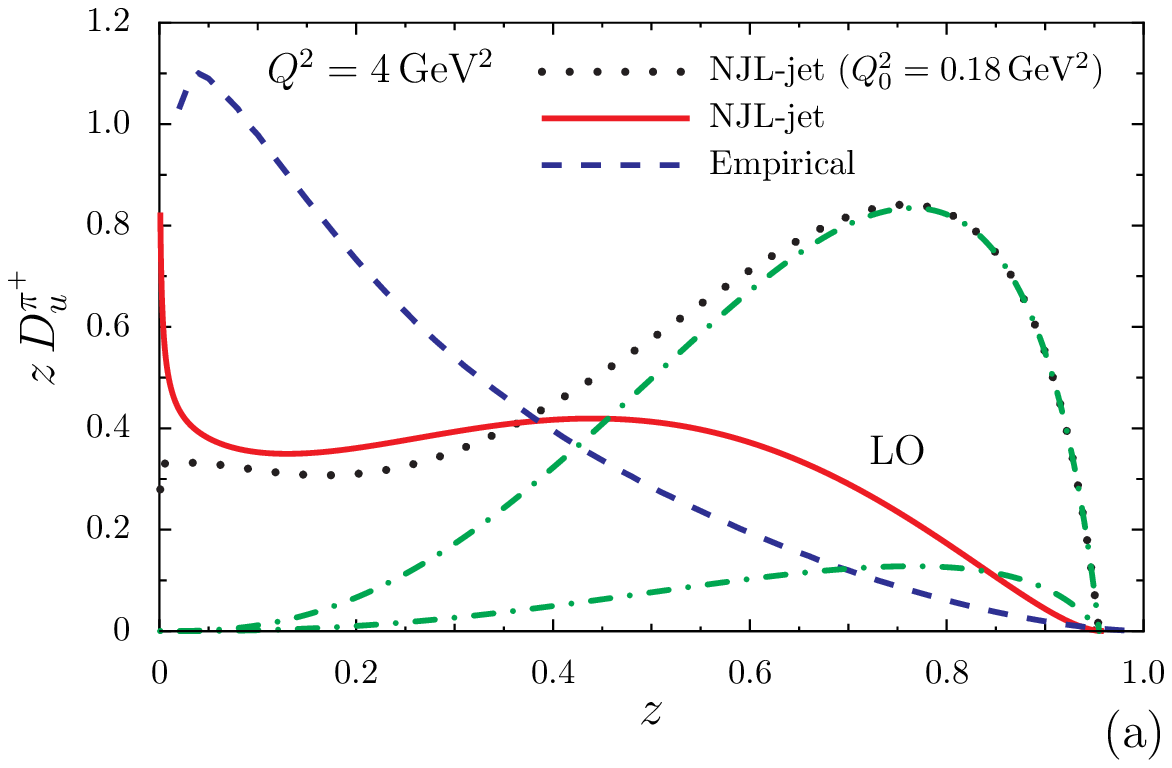}} \quad
\subfigure{\includegraphics[width=0.48\columnwidth,clip=true,angle=0]{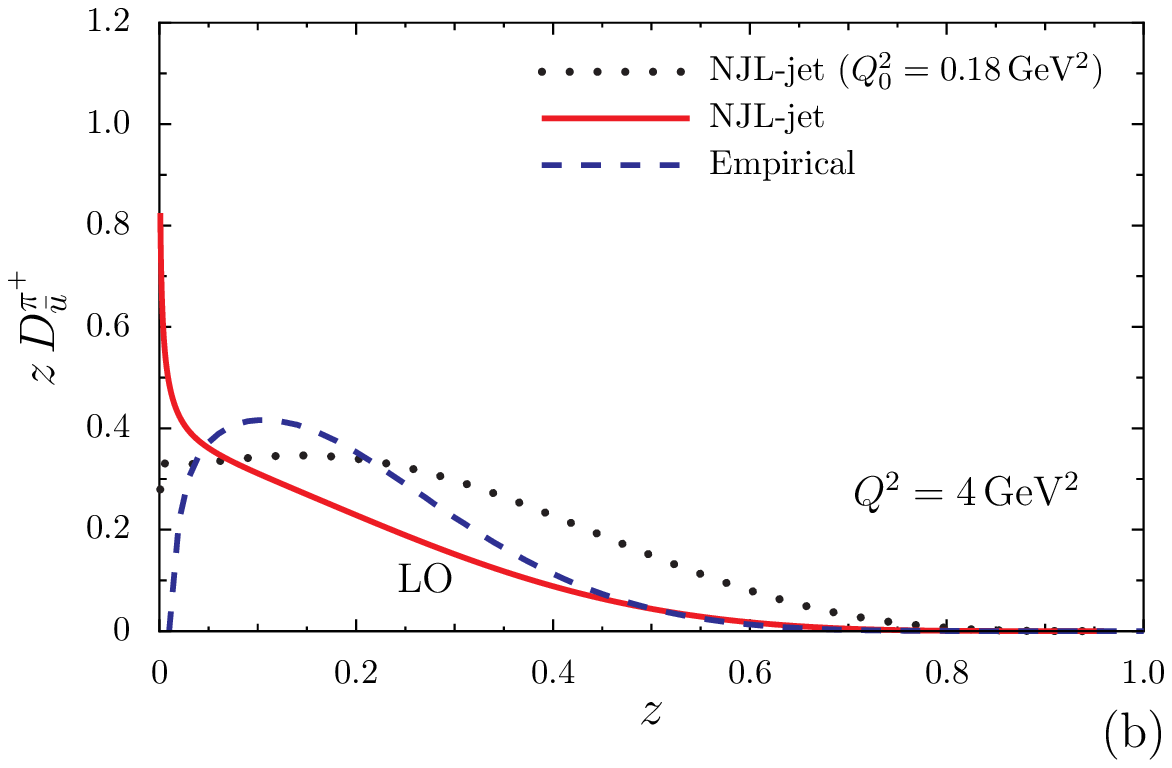}} 
\caption{
Figure (a) depicts the \textit{favoured} fragmentation function
$z D_u^{\pi^+}(z)$ and the figure (b) illustrates the \textit{unfavoured} 
fragmentation function $z D_{\bar{u}}^{\pi^+}(z)$.
In figure (a) the lower dash-dotted line is the elementary 
fragmentation function ($d^\pi_q$ of Eq.~\eqref{dqpi2}) and the upper dash-dotted
line is the renormalized elementary fragmentation function ($\hat{d}^\pi_q$
of Eq.~\eqref{relat}), which
is the driving term of the integral equation in Eq.~\eqref{final}. 
Note, these functions are zero for the unfavoured case. The solid line 
is the result after LO evolution to $Q^2=4$ GeV$^2$. The
dashed line is the empirical NLO result of Ref.~\cite{Hirai:2007cx}, evolved to $Q^2=4$ GeV$^2$.}
\label{fig:fragflavour}
\end{figure}

Fig.~\ref{fig:fragflavour}a shows the results for the favoured fragmentation function 
of Eq.~\eqref{fav} and Fig.~\ref{fig:fragflavour}b shows the unfavoured
fragmentation function of Eq.~\eqref{unfav}. 
Note, these figures correspond to half the sum and
half the difference of the curves in Figs.~\ref{fig:frag}. 
The upper dash-dotted line in Fig.~\ref{fig:fragflavour}a shows the driving 
term, $\frac{2}{3} \hat{F}(1-z)$, of the integral equation in Eq.~\eqref{final},
and the lower dash-dotted line shows the elementary fragmentation
function, $\frac{2}{3} F(1-z)$. For the unfavoured case these two
functions are zero. Both figures demonstrate the importance of cascade 
processes in the present NJL-jet model.

\section{Summary and conclusions}
\label{sec:summary}

In this paper we used the NJL model as an effective quark theory to study
the simplest fragmentation function, namely, the fragmentation of unpolarized
quarks to pions. 
Our aim was to develop a framework which satisfies the momentum and isospin sum 
rules in a natural way, without the introduction of ad hoc parameters. 
This framework should also give fragmentation functions that have the 
correct order of magnitude at intermediate and large $z$. 
We explained in detail, that for this purpose, 
the simplest approximation where a truncation is made to
the one-quark spectator state, in the defining relation given by Eq.~\eqref{ddef}, 
is completely inadequate. 
Although this approximation does not violate any conservation law, it 
gives very small fragmentation functions; because the probability for
the elementary fragmentation process is small in effective theories based on 
constituent quarks and the quark remainder can carry an appreciable 
amount of momentum.     

In order to overcome these difficulties we followed the idea of the 
quark jet-model and made a generalized product ansatz to describe the cascade 
processes in the NJL model. We explained that this ansatz 
corresponds to a binomial distribution for the number of mesons 
emitted from the quark. However, in the limit that the maximum number of
mesons becomes very large the results are independent of the form of this
distribution function. Our formulation thus represents an extension of
the original quark jet-model, which assumed an infinite number of mesons
from the outset. We have shown in detail that this NJL-jet
model describes fragmentation processes where 100\% of the initial
quark light-cone momentum is transferred to mesons. The momentum sum
rule of Eq.~\eqref{mom}, which is assumed valid in all QCD based 
empirical fits,
is then satisfied automatically without introducing any new parameters into
the theory. We have also shown that the isospin sum rule of Eq.~\eqref{iso} is 
naturally satisfied in this approach. 

The comparison with the empirical fragmentation functions shows that 
our calculated functions have the correct order of magnitude for
intermediate and large $z$. We highlighted that a straightforward extension to 
include the NLO terms in the $Q^2$ evolution
and to include the effect of primary $\rho$ and $\omega$ mesons,  as well as fragmentation 
to other hadronic channels, will improve the description.
Therefore we can conclude that our NJL-jet model provides
a reasonable framework to analyse fragmentation functions in an effective
quark theory. 

For future work in this direction it is important to derive the jet model
type product ansatz from field theory.
The rainbow-ladder scheme for the quark self-energy may provide a
suitable framework for this purpose. 
An attempt can then be made to use this
truncation scheme to consistently describe the cascade processes for the
fragmentation functions and to include the contribution from the hadron cloud 
around the quark for the distribution functions. 
However, it is important to bear in mind that
a truncation scheme which works well for fragmentation processes may
not be suitable for the distribution functions and vice versa. 
To establish a scheme which respects the sum rules and which gives a
satisfactory description of both types of processes is an important task 
for future research. 

\begin{acknowledgments}
The authors thank A. Bacchetta, H. Hirai, T.-H. Nagai, M. Stratmann, and 
K. Tanaka for helpful discussions. W.B. and T.I. 
acknowledge the hospitality of Thomas Jefferson National Lab and
Argonne National Laboratory, where part of this work was carried out
during the spring and summer of 2008. This work was supported by the Grant
in Aid for Scientific Research of the Japanese Ministry of Education,
Culture, Sports, Science and Technology, Project No. C-19540306 and
by the U.S. Department of Energy Grant No. DEFG03-97ER4014, and by the
Contract No. DE-AC05-06OR23177, under which Jefferson Science Associates,
LLC operates Jefferson Laboratory. 
\end{acknowledgments}

\appendix
\section{Proof of the DLY relation}

In this Appendix we will prove the DLY relation expressed in Eq.~\eqref{dly} in two
independent ways. First, we follow the original derivation of Ref.~\cite{Drell:1969jm} 
in terms of the hadronic tensors
and second we start from the operator definitions given in Eqs.~\eqref{fdef} and
\eqref{ddef}. In order to illustrate the spinor algebra the formulae in this
Appendix refer to the case where $h$ is a proton, however it is trivial to modify 
the expressions for the case where $h$ is a pion.    

\subsection{General crossing relations}

We consider the following Green function
\begin{align}
\overline{M}_{\beta}^a(p,p_n) = \int d^4 x \, e^{-i p \cdot x}
\langle p_n| T\left(\mathcal{O}^a(0)\overline{\Phi}_{\beta}(x)\right)|0\rangle ,
\label{gf}
\end{align}
where $\Phi_{\beta}(x)$ is an interpolating field for the
nucleon and $\mathcal{O}^a$ is another local field operator. 
We also define the $N$-amputated Green function by
$\overline{M}_{\beta}^a(p,p_n) = \overline{\Gamma}_{\gamma}^a(p,p_n)\, i G_{N, \gamma \beta}(p)$, 
where $G_N$ is the nucleon propagator. 
From the spectral representation of Eq.~\eqref{gf} or from the familiar
reduction formalism, we can derive the relations
\begin{align}
\label{m11}
\langle p_n|\mathcal{O}^a |p\rangle  &= \overline{\Gamma}^{a}(p,p_n)\sqrt{2 M_N}\,
u_N(\vect{p}s),  \\
\label{m22}
\langle\bar{p},p_n|\mathcal{O}^a|0\rangle  &= (\pm)\, \overline{\Gamma}^a(-p,p_n)
\sqrt{2 M_N} \, v_N(\vect{p}s).
\end{align}
In Eq.~\eqref{m22} the sign is ($+$) if $\mathcal{O}$ is a fermion type operator
and ($-$) if it is a boson type operator. Also, $\bar{p}$ denotes
an antinucleon with 4-momentum $p^{\mu}=(E_{N}(\vect{p}),\vect{p})$.  
The nucleon spinors are
denoted by $u_N$ and $v_N$. Our covariant normalization
implies the following matrix elements of the nucleon field operator:
$\langle 0|\Phi(0)|p\rangle  = \sqrt{2M_N}\, u_N(\vect{p}s)$ and
$\langle \bar{p}|\Phi(0)|0\rangle  = \sqrt{2M_N} \, v_N(\vect{p}s)$.
Eqs. \eqref{m11} and \eqref{m22} are the basic crossing relations which will
be used in the following.

\subsection{Comparison of hadronic tensors}

Here we use the above crossing relations to find the connection between the
hadronic tensors (spin-independent parts only)
for the processes $e h \rightarrow e' X$ and $e^+ e^- \rightarrow h X$, 
where $h$ denotes a hadron (proton)~\cite{Jaffe:1996zw}:
\begin{align}
\label{ht1} 
W_h^{\mu \nu}(p,q) &= \frac{1}{4 \pi}  \hat{\sum}_n (2\pi)^4
\delta^4(q+p-p_n)\langle p|J^{\mu}|p_n\rangle \langle p_n|J^{\nu}|p\rangle   \nonumber \\
&= \left(-g^{\mu \nu} + \frac{q^{\mu}q^{\nu}}{q^2}\right) F^h_1(x,q^2)
+ \frac{1}{p \cdot q}\left(p^{\mu} - \frac{p \cdot q}{q^2}q^{\mu}\right)
\left(p^{\nu} - \frac{p \cdot q}{q^2}q^{\nu}\right) F^h_2(x,q^2), \\
\label{ht2}
\overline{W}_h^{\mu \nu}(p,q) &= \frac{1}{4 \pi} \hat{\sum}_n (2\pi)^4
\delta^4(q-p-p_n)\langle 0|J^{\mu}|p,\overline{p_n}\rangle 
\langle p,\overline{p_n}|J^{\nu}|0\rangle   \nonumber \\
&= \left(-g^{\mu \nu} + \frac{q^{\mu}q^{\nu}}{q^2}\right) \overline{F}^h_1(z,q^2)
+ \frac{1}{p \cdot q}\left(p^{\mu} - \frac{p \cdot q}{q^2}q^{\mu}\right)
\left(p^{\nu} - \frac{p \cdot q}{q^2}q^{\nu}\right) \overline{F}^h_2(z,q^2).
\end{align}
Here $|p \rangle$ is the state of the hadron $h$ with momentum $p$ and 
we use $x = \frac{-q^2}{2 p \cdot q}$
and $z = \frac{2 p \cdot q}{q^2} = - \frac{1}{x}$. We also defined
$\hat{\sum}_n = \sum_n \int \frac{d^4 p_n}{(2\pi)^3}\,\delta(p_n^2-M_n^2)\, \Theta(p_{n0})$,
where $M_n$ is the invariant mass of the state $n$. 
Using Eq.~\eqref{m11} and its complex conjugate for the current operator $J^{\nu}$:
\begin{align}
\langle p_n|J^{\nu} |p\rangle  &= \sqrt{2M_N} \, \overline{\Gamma}^{\nu}(p,p_n) u_N(\vect{p}s),  
\label{m1a}  \\
\langle p|J^{\nu}|p_n\rangle  &= \sqrt{2M_N}\, \bar{u}_N(\vect{p}s) \Gamma^{\nu}(p,p_n), 
\label{m1b}
\end{align}
where $\Gamma^{\nu}_{\beta}= (\gamma_0 \overline{\Gamma}^{\dagger\nu})_{\beta}$, 
that is, $\overline{\Gamma}^{\nu}=\Gamma^{\nu \dagger}\gamma^0$.
We insert these relations into Eq.~\eqref{ht1}.
Since we consider the spin-independent part only, we can sum over the
nucleon spin $s$ and divide by 2, using 
$\sum_s u_N(\vect{p}s) \bar{u}_N(\vect{p}s) = \frac{\slashed{p}+M_N}{2M_N}$. This gives 
\begin{align}
4 \pi W_h^{\mu \nu}(p,q) =  \frac{1}{2}
\hat{\sum}_n (2 \pi)^4 \delta^{4}(q+p-p_n) \mathrm{Tr} 
\left[(\slashed{p}+M_N) \Gamma^{\mu}(p,p_n) \overline{\Gamma}^{\nu}(p,p_n) \right].
\label{res1}
\end{align}
For the hadronic tensor in Eq.~\eqref{ht2}, we first use charge conjugation
and then Eq.~\eqref{m22} and its complex conjugate for the current operator $J^{\mu}$:
\begin{align}
\langle 0| J^{\mu}|p,\overline{p_n}\rangle  &= \langle 0| \mathcal{C}^{-1} 
\left(\mathcal{C} J^{\mu} \mathcal{C}^{-1}\right) \mathcal{C}|p,\overline{p_n}\rangle 
=  \langle 0| \left(\mathcal{C} J^{\mu} \mathcal{C}^{-1}\right) |\overline{p},p_n\rangle ,
= -\langle 0| J^{\mu} |\overline{p},p_n\rangle   
=  \sqrt{2 M_N} \, \bar{v}_N(\vect{p}s) \Gamma^{\mu}(-p,p_n), \\
\langle p,\overline{p_n}|J^{\mu}|0\rangle  &= - \langle \overline{p},p_n|J^{\mu}|0\rangle 
=  \sqrt{2 M_N} \, \overline{\Gamma}^{\mu}(-p,p_n) v_N(\vect{p}s). 
\end{align}
We insert these relations into Eq.~\eqref{ht2},
sum over the nucleon spin $s$ and divide by 2 using 
$\sum_s v_N(\vect{p}s) \bar{v}_N(\vect{p}s) = -\frac{- \slashed{p}+M_N}{2M_N}$. 
This gives
\begin{align}
4 \pi \overline{W}_h^{\mu \nu}(p,q) = -  \frac{1}{2}
\hat{\sum}_n (2 \pi)^4 \delta^{4}(q-p-p_n) \mathrm{Tr}\left[
(-\slashed{p}+M_N) \Gamma^{\mu}(-p,p_n) \overline{\Gamma}^{\nu}(-p,p_n)
\right].
\label{res2}
\end{align}
By comparing Eqs.~\eqref{res1} with \eqref{res2} we obtain the
DLY crossing relation for the hadronic tensors:
\begin{align}
\overline{W}_h^{\mu \nu}(p,q) = - W_h^{\mu \nu}(-p,q), \qquad \text{where} \qquad
s_h=\frac{1}{2}.
\label{dlyf}
\end{align}
The minus sign in Eq.~\eqref{dlyf} comes from the Dirac algebra,
and for a spinless hadron the minus sign is changed to plus.
Eq.~\eqref{dlyf} implies the following relation between the structure 
functions in Eqs.~\eqref{ht1} and \eqref{ht2}:\footnote{By relations 
like Eq.~\eqref{sf1} we mean the following: 
Take a particular physical value of $z$ for the $(e^+, e^-)$ process 
($0<z<1$). Then the corresponding (unphysical) value of the Bjorken 
variable for the $(e,e')$ process is $x=1/z$ and Eq.~\eqref{sf1} gives
the connection between the structure functions.}    
\begin{align}
\overline{F}^h_1(z,q^2) &= - F^h_1(-x,q^2)=-F^h_1\left(\frac{1}{z},q^2\right), 
\label{sf1}\\
\overline{F}^h_2(z,q^2) &=  F^h_2(-x,q^2)= F^h_2\left(\frac{1}{z},q^2\right). 
\label{sf2}
\end{align}
The well known relation $F^h_2(x)=2x F^h_1(x)$ becomes, with 
$x\rightarrow -x$ and using the first equalities in Eqs.~\eqref{sf1} and \eqref{sf2}:
\begin{align}
\overline{F}^h_2(z) = - \frac{2}{z} \ \overline{F}^h_1(z),
\label{f12}
\end{align}
which also holds for spinless bosons.

The connection between the structure function $\overline{F}^h_1$ and
the fragmentation function $D_q^h(z)$ in the Bjorken limit is
as follows: The cross section for the process 
$e^+ e^- \rightarrow h X$ is~\cite{Jaffe:1996zw}\footnote{We remind the reader 
that the symbol $h$ denotes a particular hadron with a
specified spin direction, e.g., $p \uparrow$ (although the spin
averaged cross section considered here does not depend on the spin
direction). Therefore, the cross section measured for the case that
the spin of the produced nucleon is not observed has an additional factor
of 2, which is not included in Eq.~\eqref{c2}.}
\begin{align}
\frac{d\sigma^h}{dz} =  \frac{2 \alpha^2 \pi z}{q^2}
\left(\overline{F}^h_1(z,q^2) + \frac{z}{6}\, \,\, \overline{F}^h_2(z,q^2)\right)
=  \frac{4}{3}\, \frac{\alpha^2 \pi z}{q^2}\, \overline{F}^h_1(z).
\label{c2}
\end{align}
Usually this is divided by the total cross section for
$e^+ e^- \rightarrow$ hadrons
\begin{align}
\sigma_\mathrm{tot} = \frac{4 \pi \alpha^2}{q^2}\sum_q e_q^2 
\equiv \frac{4 \pi \alpha^2}{3 q^2} R , 
\label{crtot}
\end{align}
where $\sum_q$ refers to the quark flavour only. Then we obtain
\begin{align}
\frac{1}{\sigma_\mathrm{tot}} \frac{d\sigma^h}{dz}
=  \frac{1}{R} \, z \, \overline{F}^h_1(z).  \label{c3}
\end{align}
This is compared to the original definition of the fragmentation
function~\cite{Field:1976ve}:
\begin{align}
\frac{1}{\sigma_\mathrm{tot}} \frac{d\sigma^h}{dz}
\equiv \frac{1}{R}\,3\,\sum_q e_q^2 \,\left( D_q^h(z) + D_{\bar{q}}^h(z) \right)
\label{def}
\end{align}
to obtain
\begin{align}
\overline{F}^h_1(z) =  \frac{3}{z} \sum_q e_q^2 \left(D_q^h(z)
+ D_{\bar{q}}^h(z)\right).
\label{fd}
\end{align}
Because we know how to express
$F^h_1(x)$ in the Bjorken limit by the distribution functions $f_q^h(x)$, 
we obtain from Eq.~\eqref{sf1}:
\begin{align}
\overline{F}^h_1(z) = - F^h_1\left(\frac{1}{z}\right) = - \frac{1}{2}
\sum_q e_q^2 \, \left(f_q^h\left(\frac{1}{z}\right)
+ f_{\bar{q}}^h\left(\frac{1}{z}\right)\right). \label{fd1}
\end{align}
Comparing \eqref{fd} and \eqref{fd1} we obtain
\begin{align}
D_q^h(z) &= - \frac{z}{6} \, f_q^h\left(\frac{1}{z}\right),  
\qquad \text{where} \qquad
s_h=\frac{1}{2},
\label{fin} 
\end{align}
and a similar result holds for the antiquarks. Eq.~\eqref{fin} expresses the DLY 
relation of Eq.~\eqref{dly} between the distribution and fragmentation functions. 
For the case of a spinless hadron the minus sign in Eq.~\eqref{fin} becomes
a plus sign.

\subsection{Comparing the operator definitions}

Starting from the operator definitions given in Eqs.~\eqref{fdef} and \eqref{ddef},
we obtain
\begin{align}
f_q^h(x) &= \frac{1}{2} \hat{\sum}_n \delta \left(p_- x - p_- + p_{n-}\right)
\langle p|\overline{\psi}|p_n\rangle  \gamma^+ \langle p_n|\psi|p\rangle,   \label{fq}  \\
D_q^h(z) &= \frac{z}{6} \, \frac{1}{2}\, \hat{\sum}_n
\delta\left(\frac{p_-}{z}- p_- - p_{n-}\right) \langle p,\overline{p_n}|\overline{\psi}|0\rangle 
\gamma^+ \langle 0|\psi|p,\overline{p_n}\rangle .   
\label{opd}
\end{align}
For definiteness we consider again the case where $h$ is a proton. We use
$\mathcal{O}^a = \psi_{\alpha}$ in Eq.~\eqref{gf}, which gives
\begin{align}
\langle p_n|\psi |p\rangle  &= \overline{\Gamma}(p,p_n)\sqrt{2 M_N}\,
u_N(\vect{p}s) , \label{m1p}  \\
\langle \overline{p},p_n|\psi|0\rangle  &= \overline{\Gamma}(-p,p_n)
\sqrt{2 M_N}\, v_N(\vect{p}s). \label{m2p}
\end{align}
We insert Eq.~\eqref{m1p} and its complex conjugate into the operator 
definition, Eq.~\eqref{fq}, and average over the nucleon spin. This gives
\begin{align}
f_q^h(x) = \frac{1}{4}\hat{\sum}_n \delta\left(p_- x - p_- + p_{n-}\right)
\mathrm{Tr}\left[ (\slashed{p}+M_N) \Gamma(p,p_n) \gamma^+ 
\overline{\Gamma}(p,p_n)\right].
\label{dist}
\end{align}
For the fragmentation function in Eq.~\eqref{opd}, we use the charge 
conjugation relations of the quark field operators
$\mathcal{C} \psi_{\alpha} \mathcal{C}^{-1} = (C \gamma^0)_{\alpha \beta}
\psi_{\beta}^{\dagger}$ and 
$\mathcal{C} \overline{\psi}_{\alpha} \mathcal{C}^{-1} = \psi_{\beta} C_{\beta \alpha}$, 
where $C=i  \gamma^2 \gamma^0$, 
to rewrite the matrix elements in Eq.~\eqref{opd} as follows:
\begin{align}
\langle 0|\psi_{\alpha}|p,\overline{p_n}\rangle  &= 
(C\gamma_0)_{\alpha \beta}\langle p_n, \overline{p}|\psi_{\beta}|0\rangle ^* ,
\label{cc3} \\
\langle p,\overline{p_n}|\overline{\psi}_{\alpha}|0\rangle 
&= \langle \overline{p},p_n|\psi_{\beta}|0\rangle  C_{\beta \alpha}. \label{cc4}
\end{align}
Then we use $C \gamma^{\mu} C = (\gamma^{\mu})^T$ and Eq.~\eqref{m2p} to write 
\begin{align}
\langle p,\overline{p_n}|\overline{\psi}_{\alpha}|0\rangle  \gamma^+_{\alpha \beta}
\langle 0|\psi_{\beta}|p,\overline{p_n}\rangle 
= \bar{v}_N(\vect{p}s) \left[\Gamma(-p,p_n)\gamma^+
\overline{\Gamma}(-p,p_n)\right] v_N(\vect{p}s) \cdot 2 M_N .
\end{align}
Averaging over the nucleon spins we finally obtain
\begin{align}
D_q^h(z) = - \frac{z}{6} \frac{1}{4}\hat{\sum}_n \delta \left(\frac{p_-}{z}-p_- - p_{n-}\right)
\mathrm{Tr} \left[(-\slashed{p}+M_N)\overline{\Gamma}(-p,p_n)
\gamma^+ \Gamma(-p,p_n)\right].
 \label{frag}
\end{align}
Comparison of Eqs.~\eqref{dist} and \eqref{frag} gives
\begin{align}
D_q^h(z) = - \frac{z}{6} \, f_q^h\left(x=\frac{1}{z}\right)\bigg|_{p\rightarrow -p},  
\label{fdd}
\end{align}
where $p\rightarrow -p$ means to reverse all 4 components of $p^{\mu}$
and after this replacement $p^0=E_N(\vect{p})>0$.

We now consider the property of the distribution
function in Eq.~\eqref{dist} under $p^{\mu} \rightarrow - p^{\mu}$.
Expressing the summation $\hat{\sum}_n$ in terms of light-cone
momenta, the distribution in Eq.~\eqref{dist} can be written in the form
\begin{multline}
f_q^h(x) = \frac{1}{4} \sum_n \int \frac{d^4 k}{(2 \pi)^3}
\,\frac{\Theta(p_-(1-x))}{2 p_-(1-x)} \,
\,\delta\!\left(k_+ - e_N(\vect{p}) + e_n(\vect{p}- \vect{k})\right)\\
\delta(k_- - p_- x)\, \mathrm{Tr}\left[(\slashed{p}+M_N) \Gamma(p,p-k) \gamma^+ 
\overline{\Gamma}(p,p-k) \right], 
\label{cc}
\end{multline}
where $e_n(\vect{p}_n) = \frac{\vect{p}_{n \perp}^2 + M_n^2}{2 p_{n-}}$ 
and $e_N(\vect{p})= \frac{\vect{p}_{\perp}^2 + M_N^2}{2p_-}$.
We then replace $p^{\mu}\rightarrow - p^{\mu}$ and then  
$k^{\mu}\rightarrow - k^{\mu}$ in the integral. This gives
\begin{multline}
f_q^h(x)|_{p \rightarrow -p} = 
-\frac{1}{4} \sum_n \int \frac{d^4 k}{(2 \pi)^3}\,
\frac{\Theta(p_-(x-1))}{2 p_-(1-x)}\, \delta\!\left(k_+ - e_N(\vect{p})+ e_n
(\vect{p}- \vect{k})\right) \\
\delta(k_- - p_- x)\, \mathrm{Tr}\left[(-\slashed{p}+M_N) \Gamma(-p,-p+k) \gamma^+ 
\overline{\Gamma}(-p,-p+k)\right].
\label{cca}
\end{multline}
Because the result of taking the trace in Eq.~\eqref{cca} must be the
plus component of a Lorentz four vector constructed from $p^{\mu}$
and $k^{\mu}$, we have
\begin{align}
\mathrm{Tr}\left[(-\slashed{p}+M_N) \Gamma(-p,-p+k) \gamma^+ 
\overline{\Gamma}(-p,-p+k) \right]
= -\mathrm{Tr}\left[(\slashed{p}+M_N) \Gamma(p,p-k) \gamma^+ 
\overline{\Gamma}(p,p-k) \right].  
\label{s}
\end{align}
If we use Eq.~\eqref{cc} to define a function $F(x)$ by
$f_q^h(x) = \Theta(1-x) F(x)$, we obtain from Eqs.~\eqref{cca} and \eqref{s}: 
$f_q^h(x)|_{p \rightarrow -p} =  \Theta(x-1) F(x)$.
From Eq.~\eqref{fdd} we then obtain the connection between
the distribution and the fragmentation function as
\begin{align}
\label{rr1}
f_q^h(x) &= \Theta(1-x) F(x), \\
\label{rr2} 
D_q^h(z) &= - \Theta(1-z)\, \frac{z}{6}\, F\left(\frac{1}{z}\right).
\end{align}
Note, for spinless bosons there is no minus sign in Eq.~\eqref{rr2}.
This result agrees with Eq.~\eqref{fin} and would suggest that 
$f_q^h$ and $D_q^h$ are essentially one and the
same function, defined in different regions of the variable.

\section{DLY transformation of evolution kernels}

In this Appendix we explain the DLY based relation between the
evolution kernels for distribution and fragmentation functions, which
is known to be valid at LO~\cite{Blumlein:2000wh}. 
Using Eq.~\eqref{dly}, we consider the following transformation of the
quark and gluon distribution functions:
\begin{align}
f_q^h(x) &\rightarrow \left(\pm \frac{z}{6}\right) 
f_q^h\left(x=\frac{1}{z} \right), \\  
f_g^h(x) &\rightarrow \left(\mp \frac{z}{16}\right) 
f_g^h\left(x=\frac{1}{z}\right),
\end{align}
where the upper (lower) sign holds if $h$ is a 
boson (fermion). 
Using the well known evolution equations at LO~\cite{Ellis:1991qj}, it is easy to
derive the corresponding transformation of the evolution kernels. 
For the minus-type
(flavour non-singlet) combination $q-\bar{q}$, the 
kernel ($P_{qq}$) is unchanged. For the flavour singlet combination,
$\sum_{i=1}^{N_f} q_i + \bar{q}_i$, 
which couples to a gluon, the evolution kernel is transformed as follows:
\begin{align}
  \begin{pmatrix}
  P_{qq}(x) & P_{qg}(x) \\
  P_{gq}(x) & P_{gg}(x)
  \end{pmatrix}  \longrightarrow
  \begin{pmatrix}
  P_{qq}(z) & 2 N_f P_{gq}(z) \\
  \frac{1}{2 N_f} P_{qg}(z) & P_{gg}(z)
  \end{pmatrix}.
\end{align}
Here $N_f=3$ is the number of flavours used in the $Q^2$ evolution 
equations. For reference, we summarize the forms of the individual  
kernels below:
\begin{align}
P_{qq}(x) &= \frac{4}{3}\left[\frac{1+x^2}{(1-x)_+} + \frac{3}{2}\delta(x-1)\right], \\
P_{qg}(x) &= N_f \left[x^2 + (1-x)^2\right] ,   \\
P_{gq}(x) &= \frac{4}{3} \frac{1+(1-x)^2}{x},   \\
P_{gg}(x) &= 6 \left[\frac{x}{(1-x)_+} +   
\frac{1-x}{x} + x(1-x)\right]
+ \left(\frac{11}{2}- \frac{N_f}{3}\right) \delta(1-x).
\end{align}



\begin{thebibliography}{99}

\bibitem{Field:1976ve}
  R.~D.~Field and R.~P.~Feynman,
  Phys.\ Rev.\  D {\bf 15}, 2590 (1977);\\
  F.~E.~Close,
  ``An Introduction To Quarks And Partons'',
  \textit{Academic Press/London 1979}.

\bibitem{Altarelli:1979kv}
  G.~Altarelli, R.~K.~Ellis, G.~Martinelli and S.~Y.~Pi,
  Nucl.\ Phys.\  B {\bf 160}, 301 (1979).

\bibitem{Collins:1981uw}
  J.~C.~Collins and D.~E.~Soper,
  Nucl.\ Phys.\  B {\bf 194}, 445 (1982).

\bibitem{Jaffe:1996zw}
  R.~L.~Jaffe,
  International School of Nucleon Structure, Erice, 1995, arXiv:hep-ph/9602236.

\bibitem{Ellis:1991qj}
  R.~K.~Ellis, W.~J.~Stirling and B.~R.~Webber,
  ``QCD and collider physics'',
  Camb.\ Monogr.\ Part.\ Phys.\ Nucl.\ Phys.\ Cosmol.\  {\bf 8}, 1 (1996).

\bibitem{Barone:2001sp}
  V.~Barone, A.~Drago and P.~G.~Ratcliffe,
  Phys.\ Rept.\  {\bf 359}, 1 (2002)
  [arXiv:hep-ph/0104283].

\bibitem{Martin:2003sk}
  A.~D.~Martin, R.~G.~Roberts, W.~J.~Stirling and R.~S.~Thorne,
  Eur.\ Phys.\ J.\  C {\bf 35}, 325 (2004)
  [arXiv:hep-ph/0308087];\\
  M.~Hirai, S.~Kumano and N.~Saito  [Asymmetry Analysis Collaboration],
  Phys.\ Rev.\  D {\bf 69}, 054021 (2004)
  [arXiv:hep-ph/0312112].

\bibitem{Sutton:1991ay}
  P.~J.~Sutton, A.~D.~Martin, R.~G.~Roberts and W.~J.~Stirling,
  Phys.\ Rev.\  D {\bf 45}, 2349 (1992).

\bibitem{Cloet:2005pp}
  I.~C.~Clo\"et, W.~Bentz and A.~W.~Thomas,
  Phys.\ Lett.\  B {\bf 621}, 246 (2005)
  [arXiv:hep-ph/0504229].

\bibitem{Wakamatsu:1997en}
  M.~Wakamatsu and T.~Kubota,
  Phys.\ Rev.\  D {\bf 57}, 5755 (1998)
  [arXiv:hep-ph/9707500];\\
  M.~B.~Hecht, C.~D.~Roberts and S.~M.~Schmidt,
  Phys.\ Rev.\  C {\bf 63}, 025213 (2001)
  [arXiv:nucl-th/0008049].

\bibitem{Hirai:2007cx}
  M.~Hirai, S.~Kumano, T.~H.~Nagai and K.~Sudoh,
  Phys.\ Rev.\  D {\bf 75}, 094009 (2007)
  [arXiv:hep-ph/0702250].

\bibitem{deFlorian:2007aj}
  D.~de Florian, R.~Sassot and M.~Stratmann,
  Phys.\ Rev.\  D {\bf 75}, 114010 (2007)
  [arXiv:hep-ph/0703242];\\
  S.~Kretzer, E.~Leader and E.~Christova,
  Eur.\ Phys.\ J.\  C {\bf 22}, 269 (2001)
  [arXiv:hep-ph/0108055].

\bibitem{Ralston:1979ys}
  J.~P.~Ralston and D.~E.~Soper,
  Nucl.\ Phys.\  B {\bf 152}, 109 (1979).

\bibitem{Sivers:1989cc}
  D.~W.~Sivers,
  Phys.\ Rev.\  D {\bf 41}, 83 (1990);\\
  J.~C.~Collins,
  Nucl.\ Phys.\  B {\bf 396}, 161 (1993)
  [arXiv:hep-ph/9208213];\\
  D.~Boer, P.~J.~Mulders and F.~Pijlman,
  Nucl.\ Phys.\  B {\bf 667}, 201 (2003)
  [arXiv:hep-ph/0303034].

\bibitem{Londergan:1996vf}
  J.~T.~Londergan, A.~Pang and A.~W.~Thomas,
  Phys.\ Rev.\  D {\bf 54}, 3154 (1996)
  [arXiv:hep-ph/9604446];\\
  R.~Jakob, P.~J.~Mulders and J.~Rodrigues,
  Nucl.\ Phys.\  A {\bf 626}, 937 (1997)
  [arXiv:hep-ph/9704335];\\
  H.~Kitagawa and Y.~Sakemi,
  Prog.\ Theor.\ Phys.\  {\bf 104}, 421 (2000);\\
  J.~J.~Yang,
  Phys.\ Rev.\  D {\bf 65}, 094035 (2002);\\
  D.~Amrath, A.~Bacchetta and A.~Metz,
  Phys.\ Rev.\  D {\bf 71}, 114018 (2005)
  [arXiv:hep-ph/0504124];\\
  A.~Bacchetta, L.~P.~Gamberg, G.~R.~Goldstein and A.~Mukherjee,
  Phys.\ Lett.\  B {\bf 659}, 234 (2008)
  [arXiv:0707.3372 [hep-ph]].

\bibitem{Drell:1969jm}
  S.~D.~Drell, D.~J.~Levy and T.~M.~Yan,
  Phys.\ Rev.\  {\bf 187}, 2159 (1969);\\
  S.~D.~Drell, D.~J.~Levy and T.~M.~Yan,
  Phys.\ Rev.\  D {\bf 1}, 1035 (1970);\\
  S.~D.~Drell, D.~J.~Levy and T.~M.~Yan,
  Nucl.\ Phys.\  D {\bf 1}, 1617 (1970).

\bibitem{Blumlein:2000wh}
  J.~Blumlein, V.~Ravindran and W.~L.~van Neerven,
  Nucl.\ Phys.\  B {\bf 586}, 349 (2000)
  [arXiv:hep-ph/0004172].

\bibitem{Boros:1999zc}
  C.~Boros, J.~T.~Londergan and A.~W.~Thomas,
  Phys.\ Lett.\  B {\bf 473}, 305 (2000)
  [arXiv:hep-ph/9909413].

\bibitem{Field:1977fa}
  R.~D.~Field and R.~P.~Feynman,
  Nucl.\ Phys.\  B {\bf 136}, 1 (1978);\\
  T.~D.~Gottschalk,
  ``Hadronization And Fragmentation'',
  19th Int. School of Elementary Particle Physics, Kupari-Dubrovnik, 1983.

\bibitem{Andersson:1985qr}
  B.~Andersson, G.~Gustafson and B.~Soderberg,
  Nucl.\ Phys.\  B {\bf 264}, 29 (1986).

\bibitem{Field:1982dg}
  R.~D.~Field and S.~Wolfram,
  Nucl.\ Phys.\  B {\bf 213}, 65 (1983).

\bibitem{Bentz:1999gx}
  W.~Bentz, T.~Hama, T.~Matsuki and K.~Yazaki,
  Nucl.\ Phys.\  A {\bf 651}, 143 (1999)
  [arXiv:hep-ph/9901377].

\bibitem{Cloet:2007em}
  I.~C.~Clo\"et, W.~Bentz and A.~W.~Thomas,
  Phys.\ Lett.\  B {\bf 659}, 214 (2008)
  [arXiv:0708.3246 [hep-ph]].

\bibitem{Nambu:1961tp}
  Y.~Nambu and G.~Jona-Lasinio,
  Phys.\ Rev.\  {\bf 122}, 345 (1961);\\
  Y.~Nambu and G.~Jona-Lasinio,
  Phys.\ Rev.\  {\bf 124}, 246 (1961).

\bibitem{Collins:1992kk}
  J.~C.~Collins,
  Nucl.\ Phys.\  B {\bf 396}, 161 (1993)
  [arXiv:hep-ph/9208213].

\bibitem{Jaffe:1991ra}
  R.~L.~Jaffe and X.~D.~Ji,
  Nucl.\ Phys.\  B {\bf 375}, 527 (1992).

\bibitem{Burkardt:1995ct}
  M.~Burkardt,
  Adv.\ Nucl.\ Phys.\  {\bf 23}, 1 (1996)
  [arXiv:hep-ph/9505259].

\bibitem{Altarelli:1977zs}
  G.~Altarelli and G.~Parisi,
  Nucl.\ Phys.\  B {\bf 126}, 298 (1977).

\bibitem{Stratmann:1996hn}
  M.~Stratmann and W.~Vogelsang,
  Nucl.\ Phys.\  B {\bf 496}, 41 (1997)
  [arXiv:hep-ph/9612250].

\bibitem{Jaffe:1983hp}
  R.~L.~Jaffe,
  Nucl.\ Phys.\  B {\bf 229}, 205 (1983).

\bibitem{CL} 
  P.J. Huber,``Robust statistics'', Wiley, New York, 1981.
 
\bibitem{Lepage:1980fj}
  G.~P.~Lepage and S.~J.~Brodsky,
  Phys.\ Rev.\  D {\bf 22}, 2157 (1980).

\bibitem{Miyama:1995bd}
  M.~Miyama and S.~Kumano,
  Comput.\ Phys.\ Commun.\  {\bf 94}, 185 (1996)
  [arXiv:hep-ph/9508246].

\end{thebibliography}
\end{document}